\documentclass[aps,prl,preprint,showpacs,superscriptaddress]{revtex4-1}  

\usepackage{gensymb}
\usepackage{amssymb}
\usepackage{placeins}

\usepackage{color}
\usepackage{csquotes}
\usepackage{rotating} 
\usepackage{gensymb}
\usepackage{float}
\usepackage{esint}
\usepackage{url}
\usepackage{amssymb}
\usepackage{braket}
\usepackage{threeparttable} 
\usepackage{multirow}
\usepackage{booktabs}

\usepackage{braket}
\usepackage{array}
\usepackage{booktabs}
\usepackage{multirow}
\usepackage{url}
\usepackage{hyperref}
\hypersetup{
    unicode=true,            
    colorlinks=true,         
       citecolor    = blue,
       filecolor = blue,       
       linkcolor = blue,
       urlcolor  = blue,
}

\newcommand{\ra}[1]{\renewcommand{\arraystretch}{#1}}

\makeatletter

\newcommand{\Rmnum}[1]{\expandafter\@slowromancap\romannumeral #1@}
\makeatother

\begin{document}

\title{Stellar $^{36,38}$Ar$(n,\gamma)^{37,39}$Ar reactions and their effect on light neutron-rich nuclide synthesis}



\author{M. Tessler}
\affiliation{Racah Institute of Physics, Hebrew University, Jerusalem 91904, Israel}

\author{M. Paul} \email[Corresponding author: ]{paul@vms.huji.ac.il}
\affiliation{Racah Institute of Physics, Hebrew University, Jerusalem 91904, Israel}

 \author{S. Halfon}
 \affiliation{Soreq NRC, Yavne 81800, Israel}
 
 \author{B. S. Meyer}
 \affiliation{Department of Physics and Astronomy, Clemson University, Clemson, South Carolina 29634, USA}
 
 \author{R. Pardo}
 \affiliation{Argonne National Laboratory, Argonne, Illinois 60439, USA}
 
 \author{R. Purtschert}
 \affiliation{Physics Institute, University of Bern, 3012 Bern, Switzerland}
 
 \author{K. E. Rehm}
 \affiliation{Argonne National Laboratory, Argonne, Illinois 60439, USA}
 
 \author{R. Scott}
 \affiliation{Argonne National Laboratory, Argonne, Illinois 60439, USA}
 
 \author{M. Weigand}
 \affiliation{Goethe University Frankfurt, Frankfurt 60438, Germany}
 
 \author{L. Weissman}
 \affiliation{Soreq NRC, Yavne 81800, Israel}
 
 \author{S. Almaraz-Calderon}
 \affiliation{Argonne National Laboratory, Argonne, Illinois 60439, USA}
 
 \author{M. L. Avila}
 \affiliation{Argonne National Laboratory, Argonne, Illinois 60439, USA}
 
 \author{D. Baggenstos}
 \affiliation{Physics Institute, University of Bern, 3012 Bern, Switzerland}
 
 \author{P. Collon}
 \affiliation{Department of Physics, University of Notre Dame, Notre Dame, Indiana 46556, USA}
 
 \author{N. Hazenshprung}
 \affiliation{Soreq NRC, Yavne 81800, Israel}
 
 \author{Y. Kashiv}
 \affiliation{Department of Physics, University of Notre Dame, Notre Dame, Indiana 46556, USA}
 
 \author{D. Kijel}
 \affiliation{Soreq NRC, Yavne 81800, Israel}
 
 \author{A. Kreisel}
 \affiliation{Soreq NRC, Yavne 81800, Israel}
 
 \author{R. Reifarth}
 \affiliation{Goethe University Frankfurt, Frankfurt 60438, Germany}
 
 \author{D. Santiago-Gonzalez}
 \affiliation{Argonne National Laboratory, Argonne, Illinois 60439, USA}
 \affiliation{Department of Physics and Astronomy, Louisiana State University, Baton Rouge, Louisiana 70803, USA}
 
 \author{A. Shor}
 \affiliation{Soreq NRC, Yavne 81800, Israel}
 
 \author{I. Silverman}
 \affiliation{Soreq NRC, Yavne 81800, Israel}
 
 \author{R. Talwar}
 \affiliation{Argonne National Laboratory, Argonne, Illinois 60439, USA}
 
 \author{D. Veltum}
 \affiliation{Goethe University Frankfurt, Frankfurt 60438, Germany}
 
 \author{R. Vondrasek}
 \affiliation{Argonne National Laboratory, Argonne, Illinois 60439, USA}
   
 \date{\today}
 
 \begin{abstract}
The $^{36}$Ar$(n,\gamma)^{37}$Ar ($t_{1/2}$ = 35 d) and $^{38}$Ar$(n,\gamma)^{39}$Ar (269 y) reactions were studied for the first time with a quasi-Maxwellian
($kT \sim 47$ keV) neutron flux for Maxwellian Average Cross Section (MACS) measurements at stellar energies.
Gas samples were irradiated at the high-intensity Soreq applied research accelerator facility-liquid-lithium target neutron source and the $^{37}$Ar/$^{36}$Ar and $^{39}$Ar/$^{38}$Ar
ratios in the activated samples were determined by accelerator mass spectrometry at the ATLAS facility (Argonne National Laboratory).
The $^{37}$Ar activity was also measured by low-level counting at the University of Bern.
Experimental MACS of $^{36}$Ar and $^{38}$Ar, corrected to the standard 30 keV thermal energy, are 1.9(3) mb and 1.3(2) mb, respectively,
differing from the theoretical and evaluated values published to date by up to an order of magnitude.
The neutron capture cross sections of $^{36,38}$Ar are relevant to the stellar nucleosynthesis of light neutron-rich nuclides;
the two experimental values are shown to affect the calculated mass fraction of nuclides in the region A=36-48 during the weak $s$-process.
The new production cross sections have implications also for the use of $^{37}$Ar and $^{39}$Ar as environmental tracers in the atmosphere and hydrosphere.
\end{abstract}


\maketitle

The argon isotopes $^{36}$Ar and $^{38}$Ar are among the rare stable nuclides for which no experimental neutron-capture cross sections exist above thermal energy.
While the abundances of $^{36,38}$Ar in terrestrial atmospheric argon are very low relative to $^{40}$Ar (produced mainly from $^{40}$K decay \cite{von, Anders}),
$^{36}$Ar (84.59\%) and $^{38}$Ar (15.38\%) are the major argon isotopes in the solar system \cite{Lodders} and likely so in stellar matter.
They are expected, together with the branching point $^{39}$Ar, to play a role in nucleosynthesis of light neutron-rich nuclei (\textit{e.g.} $^{36}$S, $^{40}$Ar, $^{40}$K),
believed to be produced during the weak $s$-process phase of stellar evolution \cite{Hoffman, Reifarth}.
The $^{40}$K ($t_{1/2}$=1.248(3) Gy \cite{40K}) nuclide, in particular, is an important cosmo- or geochronometer and was used  to estimate the age and duration of the
$s$-process as  $\sim$10 Gy \cite{B2FH,Beer_Pen}.
$^{40}$K can be produced also in explosive oxygen burning \cite{Clayton} as a primary nucleosynthesis product in a massive star of initially pure hydrogen while 
the (secondary) $s$-process production of $^{40}$K requires initial abundances of heavy species. A better understanding of Ar cross sections 
will help clarify the relative primary vs. secondary production of $^{40}$K.
In a different realm of study, the half-life of $^{37}$Ar ($t_{1/2}$=35.011(19) d \cite{NDS_37}) makes this isotope an ideal chronometer for studying
circulation and mixing \cite{Loosli1}, and that of $^{39}$Ar (269(3) y \cite{39Ar_t})
for dating groundwater \cite{Corcho, Loosli2} and ocean water up to about 1000 years \cite{Schlosser}.
The atmospheric steady state concentrations of $^{37}$Ar and $^{39}$Ar are mainly determined by the spallation reactions $^{40}$Ar$(n,4n)^{37}$Ar and $^{40}$Ar$(n,2n)^{39}$Ar
and at lower neutron energies by the $^{36}$Ar$(n,\gamma)^{37}$Ar and $^{38}$Ar$(n,\gamma)^{39}$Ar reactions \cite{Loosli1}.
The latter are also relevant for the estimation of anthropogenic emissions from nuclear installations or for nuclear explosion monitoring \cite{LLC2}.

We measured the $^{36}$Ar and $^{38}$Ar neutron capture cross sections by activation with quasi-Maxwellian neutrons produced by the $^7$Li$(p,n)$ reaction at the
superconducting linear accelerator of Soreq applied research accelerator facility (SARAF) \cite{SARAF1, SARAF2} and the Liquid-Lithium Target (LiLiT) \cite{LiLiT1,LiLiT2}.
The activation products $^{37}$Ar and $^{39}$Ar were counted offline by accelerator mass spectrometry (AMS);
$^{37}$Ar production was also determined by Low-Level Counting (LLC).
Neutron irradiation of separate $^{36}$Ar and $^{38}$Ar samples was performed at the pneumatic transfer tube (rabbit) of the Soreq IRR-1
nuclear reactor in order to re-measure the respective thermal neutron capture cross sections.
Preliminary results of these experiments were reported in \cite{INPC16, AMS14}.

\begin{figure}
\centering
\includegraphics[width=0.99\columnwidth]{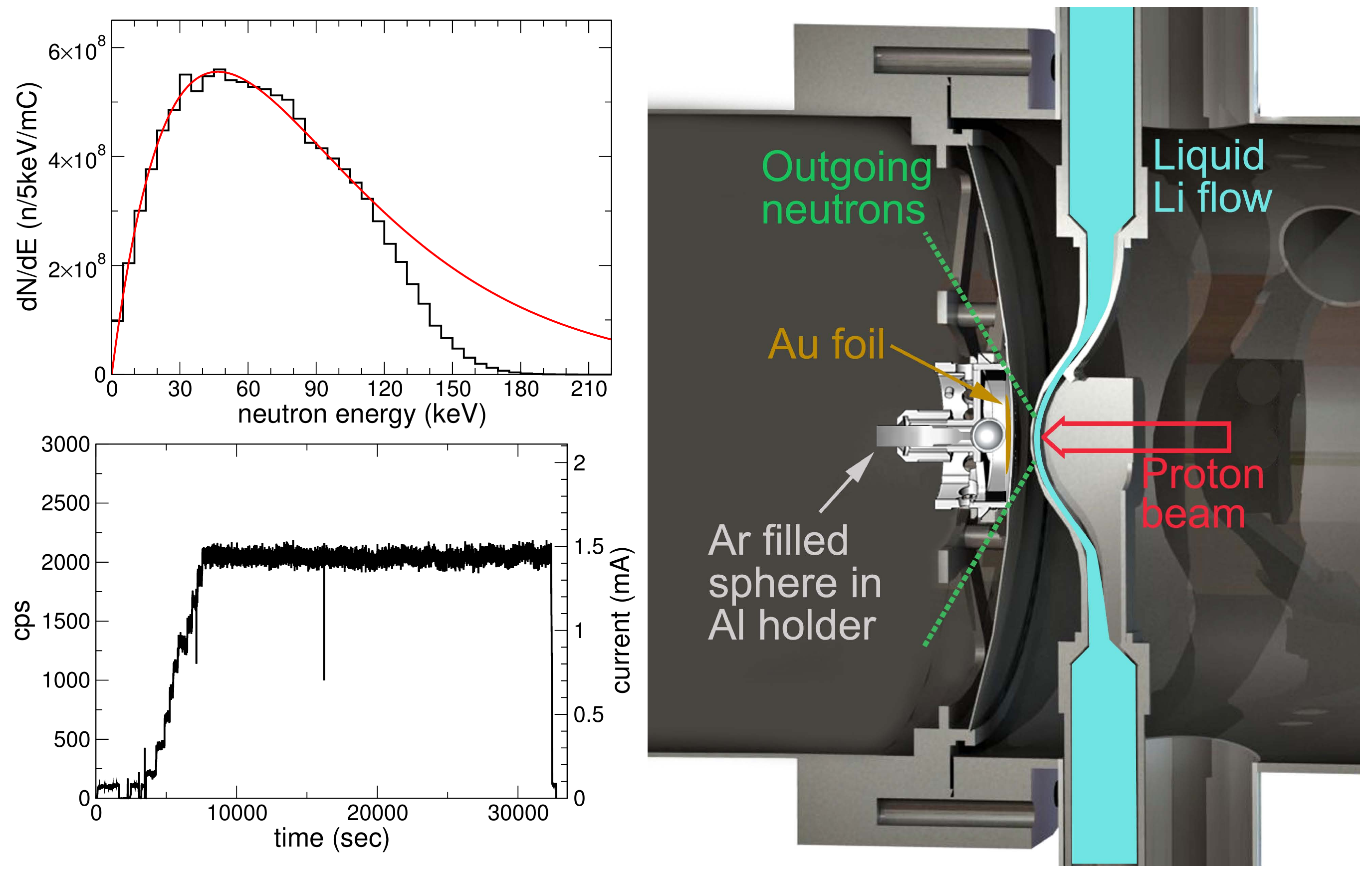}
  \caption{\label{fig:ex_setup} (Color online) (right) Diagram of the Liquid-Lithium Target (LiLiT) and activation target assembly.
  The ($\sim$1.5 mA, $\sim$9 mm full width) proton beam (open red arrow) impinges on the free-surface lithium film (cyan) (see \cite{LiLiT1} for details).
  The Ar-filled sphere and Au foil are positioned in the outgoing neutron cone (green dotted lines)
  in a vacuum chamber separated from the LiLiT chamber.
  (top left) Simulated neutron spectrum incident on the $^{36}$Ar sample (black) and a fit in the range $E_n\sim0-110$ keV with a Maxwell-Boltzmann flux (red) at $kT \sim 47$ keV.
  (bottom left) Count rate (left y-axis) of fission chamber (see text) and calibration to proton current (right y-axis) during the $^{36}$Ar run.}
\end{figure}

Enriched $^{36}$Ar, $^{38}$Ar and mixed $^{38}$Ar+$^{nat}$Ar gas samples were filled 
into Ti spheres (10 mm outer diameter, 0.2 mm thick Ti shell) \cite{sphere}.
Due to the thermodynamical properties of Ar, the filling was made 
by successive compression with a custom-made piston and cryogenic pumping in order to achieve the required pressure ($\sim$30 bar).
The samples used are listed in Table \ref{table:samples}.
\begin{table}
\centering
\caption{\label{table:samples}Samples used and the results of the $^{A+1}$Ar/$^A$Ar ratios.
$^{36}$Ar and $^{38}$Ar gas samples \cite{enrich} were enriched to 99.935\% and 99.957\% for the respective isotopes.
The final $^{37}$Ar/$^{36}$Ar ratios were obtained by taking a weighted average of the AMS and LLC results (Fig. \ref{fig:AMS_LLC}).
Sphere \#52a was irradiated with 1 mm thick Cd shield to estimate the epithermal neutron fraction.
The $^{38}$Ar/$^{nat}$Ar ratio for sphere 54 (52b) is 11.7 (10.2). For more details see the Supplemental Material \cite{Supp}.}
\begin{ruledtabular}
\ra{1.3}
\begin{tabular}{l c c c c}
Sphere \# & $^A$Ar (mg)& $^{A+1}$Ar/$^A$Ar ratio\\ [0.5ex]
39 (LiLiT) & $^{36}$Ar (24.5)  & $8.6(6)\times10^{-13}$\\
52a (reactor, Cd) & $^{36}$Ar (19.9)& $1.4(1)\times10^{-12}$\\
60 (reactor) & $^{36}$Ar (22.6)& $3.3(2)\times10^{-10}$\\ [0.5ex]
59 (LiLiT) & $^{38}$Ar (19.5) & $4.0(4)\times10^{-13}$\\
54 (reactor) & $^{38,nat}$Ar (12.8) & $8.6(9)\times10^{-11}$\\
52b (reactor) & $^{38,nat}$Ar (15.8) & $1.8(2)\times10^{-11}$\\ [0.5ex]
\end{tabular}
\end{ruledtabular}
\end{table}
For the samples irradiated at SARAF-LiLiT (Table \ref{table:samples}), each gas sphere was placed with a 25 mm-diameter Au foil (12.5 $\mu$m thick),
used as a neutron fluence monitor in an evacuated chamber downstream of LiLiT (Fig. \ref{fig:ex_setup}).
LiLiT consists of a windowless film of liquid lithium (1.5 mm thick, 18 mm wide) flowing at 2-3 m/s, serving as both
the neutron-producing target and the kW-power beam dump for the incident $\sim$1.5 mA proton beam \cite{LiLiT1, LiLiT2}.
The distance from the neutron source to the center of the Ar-filled sphere was 11.3 mm, intercepting $\sim$ 30\% of the outgoing neutrons.
The proton beam energy, measured by Rutherford back scattering off a Au target after the acceleration module,
was found to be $1932\pm3$ ($1940\pm3$) keV for the $^{36}$Ar ($^{38}$Ar) irradiation.
A proton beam energy spread of $\sim$15 keV, estimated from beam dynamics calculations,
was verified experimentally \cite{gitai_thesis}.
Auto-radiographic scans \cite{nic13} of the Au foils were conducted to determine proton beam centering.
An offset of 2.0 (2.2) mm for the $^{36}$Ar ($^{38}$Ar) irradiation was found; this offset was accounted for in our simulations.
The neutron yield was continuously monitored with a fission-product ionization chamber \cite{FC}, located $\sim$80 cm downstream the target at 0$\degree$.
The fission chamber count rate was calibrated to beam current (at low intensity) using a Faraday cup located $\sim$1 m upstream of the Li target.
The total integrated current was $\sim$10.8 (7.35) milliampere hour for the $^{36}$Ar ($^{38}$Ar) irradiation (Fig. \ref{fig:ex_setup}).

The $^{37}$Ar nuclide decays by pure electron capture with no $\gamma$-ray emission; $^{37}$Ar is notable for its role in Davis' solar neutrino experiment \cite{Davis} where its production via
$^{37}$Cl$(\nu_e,e^-)^{37}$Ar was detected by Auger electron counting.
We detected and counted for the first time $^{37}$Ar by Accelerator Mass Spectrometry (AMS) at the ATLAS facility of Argonne National Laboratory to measure the $^{37}$Ar/$^{36}$Ar
ratio of the irradiated samples. Ar gas was directly fed from the sphere container into
an Electron Cyclotron Resonance (ECR) ion source through a remote-controlled sapphire leak valve.
$^{36,37}$Ar$^{8+}$ ions were extracted from the ion source and accelerated alternately through ATLAS at an energy of 6 MeV/$u$ by appropriate scaling of all accelerator elements.
It was found necessary to strip the $^{37}$Ar$^{8+}$ ions and count $^{37}$Ar$^{18+}$ (fully stripped) in order to suppress the $^{37}$Cl (Z=17) background.
Stripping was done with a 200 $\mu$g/cm$^2$ C foil at an intermediate stage of the ATLAS linear accelerator. The stripping process (normally not used in AMS measurements at ATLAS)
however produced an isotope fractionation and the effective beam transmission efficiency ($1.84(18)\times10^{-2}$) was determined by interpolation between the measured $^{36}$Ar and $^{38}$Ar transmissions.
The $^{37}$Ar$^{18+}$ ions were counted using a $\Delta$E-E telescope of Si detectors, 50 and 300 $\mu$m thick, respectively, showing background-free spectra;
the detection sensitivity in the present experiment was $^{37}$Ar/Ar $\sim10^{-15}$ (see Supplemental Material \cite{Supp}).

\begin{figure}[h]
\centering
\includegraphics[width=0.8\columnwidth]{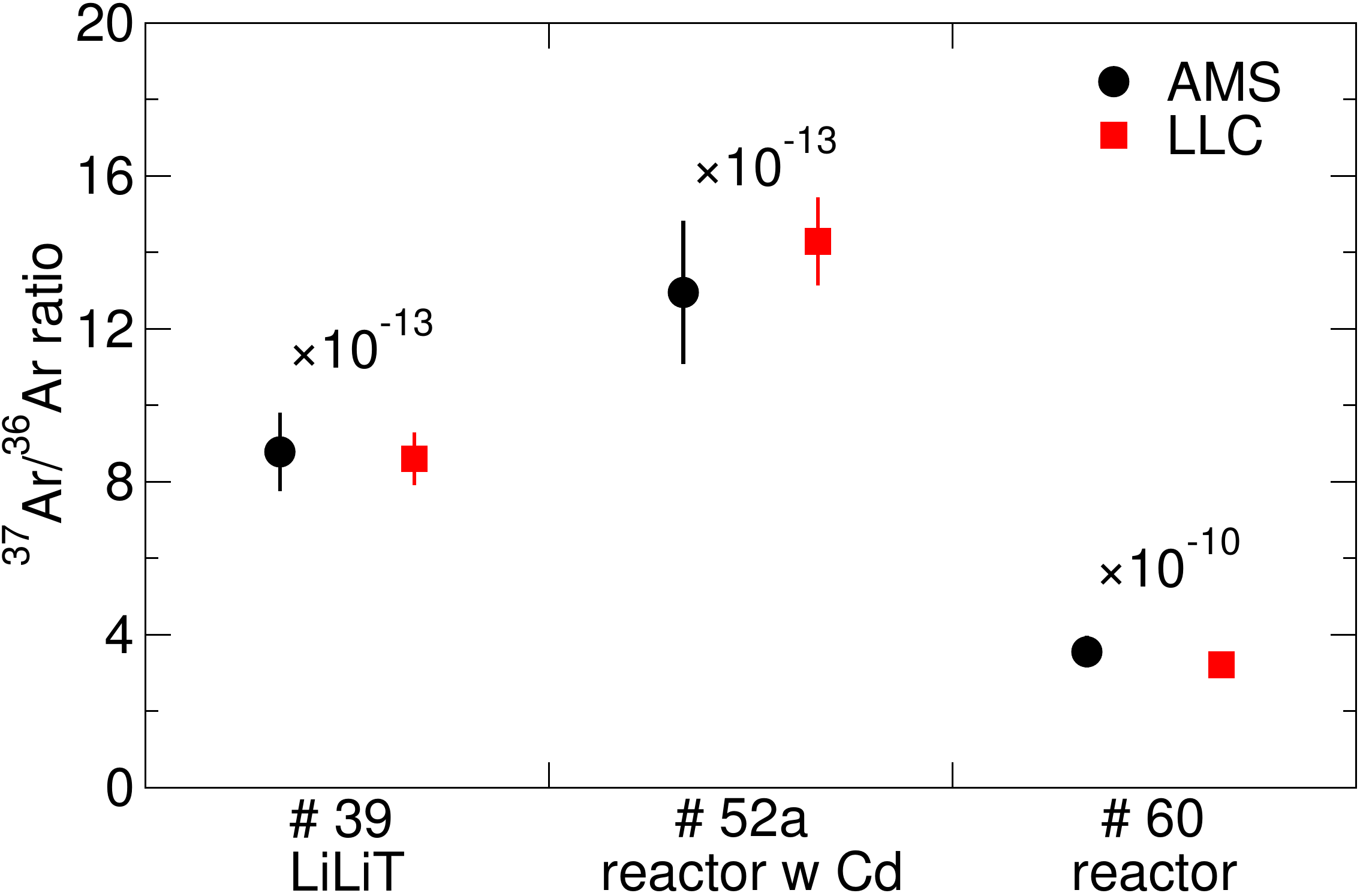}
  \caption{\label{fig:AMS_LLC} (Color online) Comparison of the $^{37}$Ar/$^{36}$Ar ratio (at the end of irradiation) measured by AMS (black) and LLC (red).}
\end{figure}

The $^{37}$Ar activity of the same samples was also determined by ultra-low-level counting (LLC) in a second stage.
Stainless steel vials containing $\sim$1 cm$^3$ aliquots of the same activated samples were shipped to the University of Bern.
Each gas was quantitatively transferred into a 100 cm$^3$ copper proportional counter which was then filled with P6 gas (6\% methane + 94\% commercial $^{37}$Ar-free argon) to a pressure of $\sim$6 bars.
The $^{37}$Ar activity was measured by detecting Auger electrons in an underground LLC laboratory during 1-2 days \cite{LLC1, LLC2}.
Energy calibration was performed with copper K-shell X-rays (E=8.133 keV) induced by an external $^{241}$Am $\gamma$ source.
The $^{37}$Ar peak was identified at the K-capture decay energy of 2.82 keV \cite{LLC3} and integrated by means of a Gaussian fit \cite{Supp}.
The amount of $^{36}$Ar in the sample was determined, after $^{37}$Ar counting, from the filling pressure of the detector and the $^{40}$Ar/$^{36}$Ar ratio measured by mass spectrometry \cite{Finn}
using established procedures.
The overall uncertainty of 8\% of the final $^{37}$Ar/$^{36}$Ar ratio is dominated by counting statistics and the uncertainties of counting yield (5\%) \cite{Supp}.
A comparison of the $^{37}$Ar/$^{36}$Ar ratios measured by AMS and LLC is illustrated in Fig. \ref{fig:AMS_LLC}.

\begin{figure}[h]
\centering
\includegraphics[width=0.8\columnwidth]{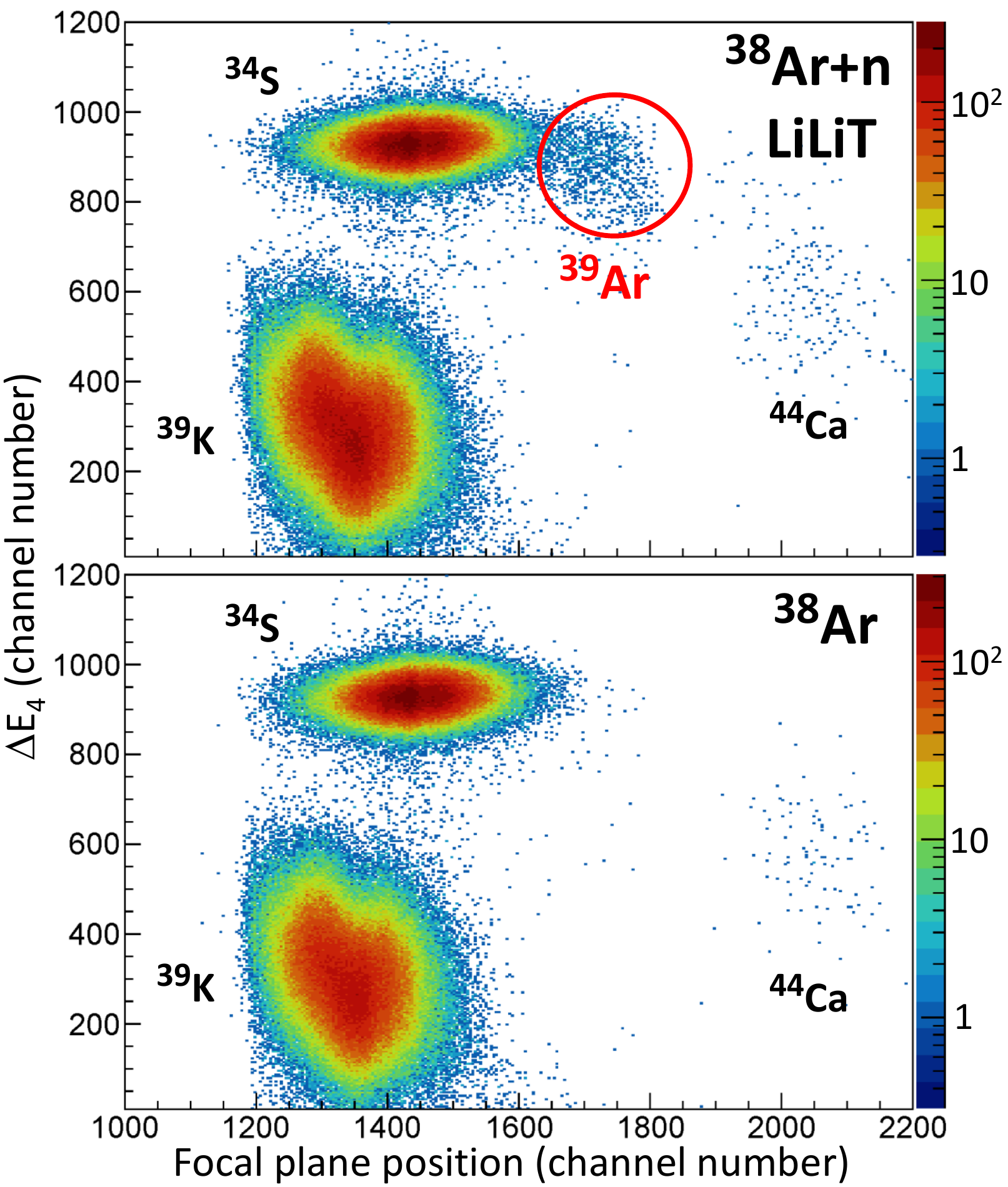}
  \caption{\label{fig:39Ar} Identification spectrum of $^{39}$Ar ions in the detector measured for the LiLiT irradiated $^{38}$Ar gas (top) and for non-irradiated $^{38}$Ar gas (bottom).
The horizontal axis represents dispersion along the focal plane and the vertical axis a differential energy loss signal measured in the fourth anode of the focal-plane ionization chamber \cite{Paul}.}
\end{figure}

Accelerator Mass Spectrometry of $^{39}$Ar has been previously performed at ATLAS \cite{Ar39_ANL1, Ar39_ANL2}.
A high ion energy is essential for the separation and discrimination of $^{39}$Ar from the extremely intense
source background of the stable $^{39}$K isobar. In our experiment, the ECR was operated at low power to reduce as much as possible impinging of the plasma onto the chamber walls,
believed to be a source of $^{39}$K contamination. $^{38,39,40}$Ar$^{8+}$ ions were accelerated to 6 MeV/$u$, similarly as described before and $^{39}$Ar$^{8+}$ ions were analyzed in the Enge gas-filled
magnetic spectrograph \cite{GFM}, which physically separates $^{39}$Ar from beam contaminants, \textit{e.g.} $^{39}$K$^{8+}$ and $^{34}$S$^{7+}$, which have close-by m/q values (Fig. \ref{fig:39Ar}).
The accelerator transmission efficiency for $^{39}$Ar$^{8+}$ (0.40(3)) was interpolated between those of $^{38}$Ar$^{8+}$ and $^{40}$Ar$^{8+}$ \cite{Supp}.

The ratios $r{\ }={\ }^{A+1}$Ar/$^{A}$Ar at the end of irradiation are determined by $r = \frac{N_{A+1}}{\epsilon {\ }t}\frac{q e}{10^{-9}{\ } i_{A}} e^{\lambda t_{cool}}$
where $N_{A+1}$ is the number of $^{A+1}$Ar detected, $\epsilon$ is the detector efficiency (measured to be 0.91(3) for $^{38}$Ar due to grid shadowing
in the spectrograph focal-plane detector), $t$ the counting time,
$q$ is the ion charge state (18 for $^{37}$Ar and 8 for $^{39}$Ar), $e$ is the electronic charge in coulomb, and $i_{A}$ the $^A$Ar$^{q+}$ beam intensity (nanoampere);
$\lambda=\frac{ln(2)}{t_{1/2}}$ is the $^{A+1}$Ar decay constant and $t_{cool}$ is the time between the end of irradiation and counting.
The final results of the $^{A+1}$Ar/$^A$Ar ratios for all gas samples are presented in Table \ref{table:samples}.

In the reactor irradiations, two small Au samples were attached to the $^{36}$Ar and $^{38}$Ar spheres for neutron monitoring, using 98.65(9) b \cite{Atlas} for the $^{197}$Au thermal
neutron capture cross section. A minor correction for the epithermal activation of Au was applied, using the $^{198}$Au activity measured for a gas sphere entirely shielded with 1 mm thick Cd.
In contrast to the $^{36}$Ar sample, two $^{38}$Ar samples irradiated at the reactor (Table \ref{table:samples}) were mixed with $^{nat}$Ar to use $^{41}$Ar
($\sigma_{th}(^{40}$Ar)=0.66(1) b \cite{Atlas}) as an internal neutron monitor in addition to the Au monitors; excellent agreement was obtained between the two neutron fluence
calibrations \cite{Supp}.
The $^{36,38}$Ar measured thermal capture cross sections are listed in Table \ref{table:MACS_comp}. Uncertainties (1$\sigma$) for $^{36}$Ar ($^{38}$Ar) are 3\% (2\%) and 7\% (11\%)
from the neutron fluence and atom ratio determinations, respectively.

For the LiLiT irradiated samples, the average experimental cross section, $\sigma_{exp}$, is obtained by $\sigma_{exp} = \frac{r}{\Phi_{n}}$,
where $\Phi_n$ is the effective neutron fluence (n/cm$^2$). In view of the complex geometry of the gas sphere irradiation, $\Phi_n$ is calculated as $\Phi_n = \frac{\sum l_n}{V}$ where $V$ (0.46 cm$^3$)
is the gas sphere's volume, $l_n$ is the length a neutron travels inside the Ar gas and $\sum l_n$ is the sum of the lengths traveled by all the neutrons inside the Ar gas sphere during the irradiation.
$\sum l_n$ is calculated by a detailed simulation (see below), taking a statistically representative sample of neutrons and scaling by the Au activity.
The validity of the expression $\frac{\sum l_n}{V}$ for the neutron fluence, $\Phi_n$, was confirmed by comparing the value 
calculated in this way for 
the Au (planar) monitor
with its measured activity; experimental and calculated values agree within 0.5\%.
The values of $\Phi_n$ (n/cm$^2$) and $\sigma_{exp}$ for $^{36}$Ar ($^{38}$Ar) are 6.2(1)$\times10^{14}$ (4.22(9)$\times10^{14}$) and 1.4(1) mb (0.95(10) mb), respectively.
Uncertainties (1$\sigma$) for the $^{36}$Ar ($^{38}$Ar) experimental cross section $\sigma_{exp}$ are 2\% (2\%) and 7\% (11\%) from the neutron fluence and atom ratio determinations, respectively.

The experimental cross section measured in our experiments is an energy-averaged value over the neutron spectrum and interpretation in terms of a
Maxwellian Averaged Cross Section (MACS) requires knowledge of the shape of the spectrum.
The integral neutron spectrum seen by the targets under the irradiation conditions of the experiment is however not measurable. Instead we rely
on detailed simulations using the codes SimLiT \cite{SimLiT} for the thick-target $^7$Li$(p,n)$ neutron yield, and GEANT4 \cite{GEANT4}
for neutron transport (Fig. \ref{fig:ex_setup}) \cite{PLB_Zr}. The SimLiT-GEANT4 simulations have been carefully benchmarked in separate experiments
and excellent agreement with experimental time-of-flight and (differential and integral) energy spectra was obtained \cite{PLB_Zr,SimLiT,Gitai}.
The simulated neutron spectrum,$\frac{dn_{sim}}{dE_n}$ is well fitted in the range $E_n\sim0-110$ keV ($\sim$ 90\% of the incident neutrons) by a Maxwell-Boltzmann (MB) flux
$v\frac{dn_{MB}}{dE_n} \propto E_n exp(-E_n/kT)$ with $kT\sim47$ keV (Fig. \ref{fig:ex_setup}).
The quantitative normalization of the neutron spectrum, $\frac{dn_{sim}}{dE_n}$, was obtained by comparing the experimental number of $^{198}$Au nuclei
(measured by gamma activity with a high-purity germanium detector) in the Au foil monitor with the number of $^{198}$Au nuclei calculated in the detailed simulation of the entire setup (see \cite{PLB_Zr} for details).

We calculate the MACS at a given thermal energy $kT$ with the procedure developed in 
\cite{PLB_Zr,PLB_Zr2}, using the expression $MACS(kT) = \frac{2}{\sqrt\pi} C_{H\mbox{-}F}(kT)  \sigma_{exp}$
where the correction factor $C_{H\mbox{-}F}(kT)$ is given by
\begin{equation}
C_{H\mbox{-}F}(kT) = \frac{\int_0 ^{\infty} \sigma(E_n) E_n e^{-\frac{E_n}{kT}}dE_n}{\int_0 ^{\infty} E_n e^{-\frac{E_n}{kT}}dE_n} / \frac{\int_0 ^{\infty} \sigma(E_n) \frac{dn_{sim}}{dE_n} dE_n}{\int_0 ^{\infty} \frac{dn_{sim}}{dE_n} dE_n}. \label{eq:C} 
\end{equation}
$\sigma(E_n)$ may have coherent contributions from compound-resonances and (weakly energy dependent) direct captures (DC). We note here that $\sigma_{exp}$ includes all contributions
in the experimental energy range; we use in Eq. (\ref{eq:C}) the Hauser-Feshbach model for the energy dependence of $\sigma(E_n)$ in the wider MB range and estimate the additional uncertainties associated with direct capture.
In order to account for the sensitivity to the low density of available compound states in $^{37,39}$Ar, we apply different codes \cite{Supp}: 
TENDL-2014 \cite{TENDL14}, -2015 \cite{TENDL15}, -2017 \cite{TENDL17} and TALYS-1.8 \cite{TALYS}
with a microscopic level density and average the $C_{H\mbox{-}F}(kT)$ values obtained; the greater of 20\% of the correction or their standard deviation is attributed to the MACS corrections. 
It should however be noted that the extrapolation of the MACS to different thermal energies and determination of their uncertainties were made using a limited number of theoretical models,
due to the total absence of experimental knowledge of resonances in the $^{37,39}$Ar compound nuclei.
We also add an estimated independent 15\% uncertainty from s-wave and p-wave DC contributions.
Detailed calculations of the correction factor and its uncertainties will be included in an expanded version of this Letter.
Our MACS values and uncertainties are listed in Table \ref{table:MACS_comp} and compared to existing theoretical values.
\begin{table}
\centering
\caption{\label{table:MACS_comp}Comparison of the experimental thermal cross sections and  MACS(30 keV) obtained in this work to theoretical and evaluated data.
}
\begin{ruledtabular}
\ra{1.3}
\begin{tabular}{l c c}
Year [Ref.] & $^{36}$Ar  & $^{38}$Ar\\ [0.5ex]
& \multicolumn{2}{c}{thermal cross section (b)}\\
 1950 \cite{McMurtie} & 6.5(10)\\
 1968 \cite{Wille} & 5.0(8)\\
 1952 \cite{Katcoff} &  & 0.8(2)\\
 2006 \cite{Atlas} & 5.2(5) & 0.8(2) \\
 This work & 3.9(3) & 0.68(8)\\[0.5ex]
 \hline
&  \multicolumn{2}{c}{MACS(30 keV) (mb)}\\

1978 \cite{Woosley} & 6.7 & 2.6\\
1983 \cite{Grup} & 8  \\
2000 \cite{Rauscher} & 14 & 3.9 \\
2005 \cite{Goriely} & 24.6 & 8.07 \\
2011 \cite{ENDF, Jeff, Rosfond} & 8.86 & 0.137 \\
2015 \cite{TENDL15} & 8.48 & 2.82 \\
Kadonis \cite{Kadonis} & 9.0(15) & 3.0(3) \\
This work (30 keV) & 1.9(3) &  1.3(2) \\
This work (47 keV) & 1.4(2) & 0.92(16) \\ [0.5ex]
\end{tabular}
\end{ruledtabular}
\end{table}

\begin{figure}[h]
\centering
\includegraphics[width=0.8\columnwidth]{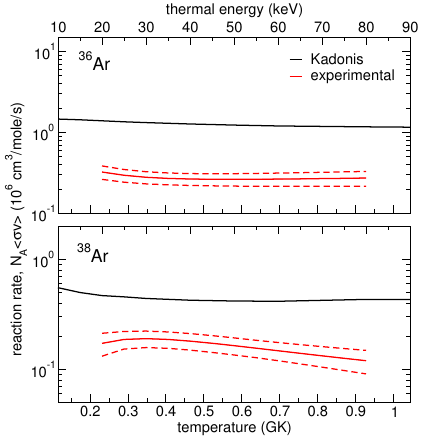}
  \caption{\label{fig:reac_rate} (Color online) Comparison of the $^{36}$Ar (top) and $^{38}$Ar (bottom) $(n,\gamma)$ reaction rates ($N_A\left<\sigma v\right>$) extracted from this work (red)
  to the Kadonis \cite{Kadonis} recommended values (black). The dashed curves encompass the estimated 1$\sigma$ uncertainty.}
\end{figure}

\begin{figure}[h]
\centering
\includegraphics[width=0.8\columnwidth]{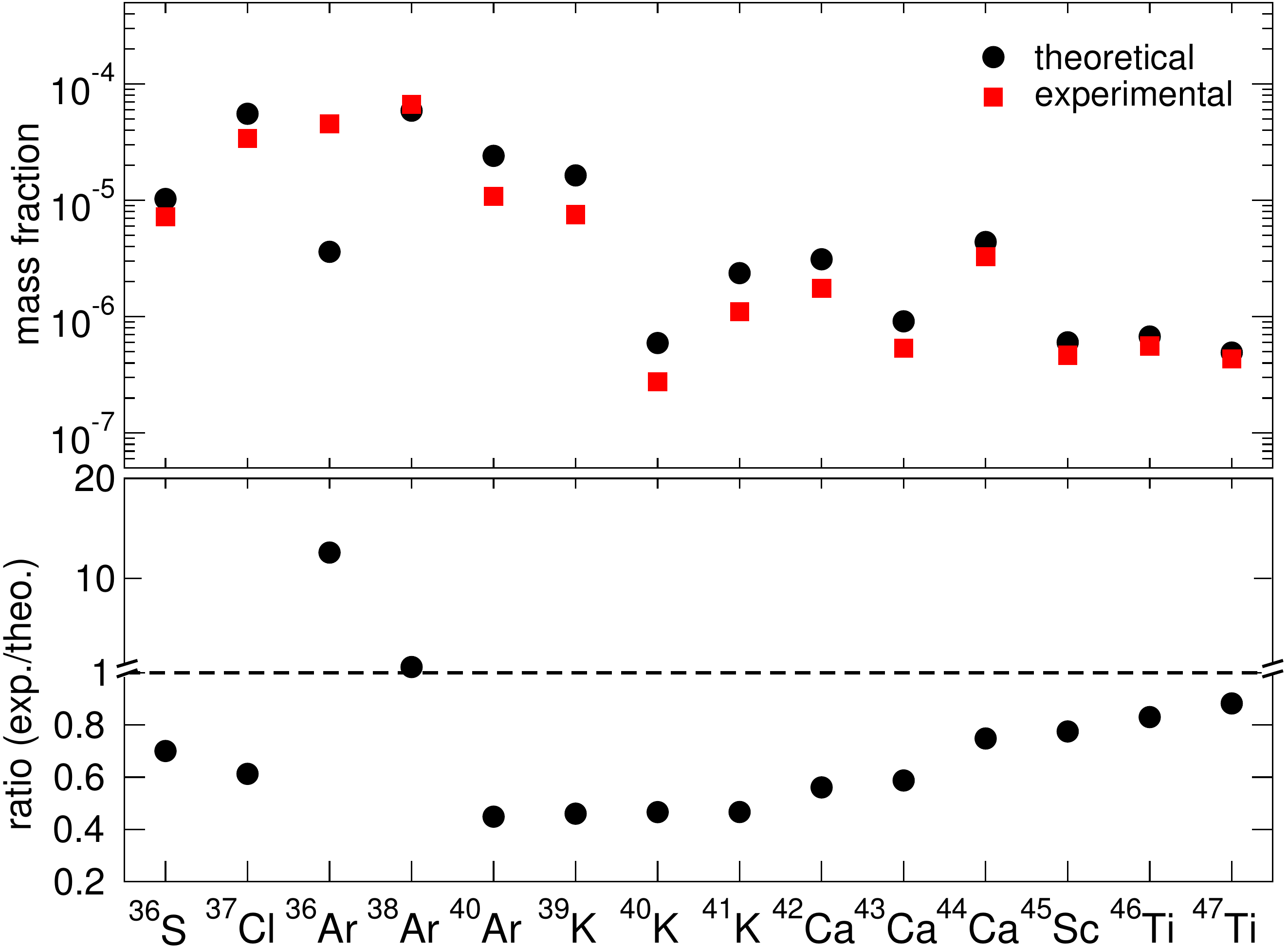}
  \caption{\label{fig:mass_frac_comp}(top) Comparison of the mass fractions calculated for stable nuclei between $^{34}$S and $^{58}$Fe changing by $>$10\% at the end of a single-zone
  calculation modeling He burning in a massive star, using literature rates \cite{Kadonis} (solid circles) or replacing the $^{36,38}$Ar$(n,\gamma)$ rates with the experimental values from this work
  (solid squares). We observe a smoother distribution of mass fractions in the vicinity of $^{36}$Ar when using the experimental cross sections. (bottom) Ratio of the mass fractions using 
  experimental and literature reaction rates as above.}
\end{figure}

The experimental MACS values (Table \ref{table:MACS_comp}) obtained in this work are notably different from previous calculations.
Fig. \ref{fig:reac_rate} shows the $^{36,38}$Ar$(n,\gamma)$ reaction rates ($N_A\left<\sigma v\right>$) based on our measurements and extrapolation to different
temperatures, compared to the rates adopted so far \cite{Kadonis}.
In order to show the potential effect of these experimental rates on stellar nucleosynthesis, we performed a single-zone network calculation using physical conditions appropriate
for the He core burning phase of a massive star in which the new $^{36,38}$Ar$(n,\gamma)$ rates are used, leaving all others unchanged \cite{Kadonis}.
The calculations are done using the single-zone NucNet Tools reaction network code \cite{nucnet} starting at the H-burning phase with solar abundances \cite{Lodders} and continuing into a single-zone
He core burning (T= 300 MK, density of 1 kg/cm$^3$).
Substantial (10-50\%) changes in the calculated mass fractions for neutron-rich light nuclides between $^{34}$S and $^{58}$Fe are observed (Fig. \ref{fig:mass_frac_comp}),
reminiscent of the sensitivity observed in the weak $s$-process region (A$\sim$56-70) due to the change of a single cross section \cite{Nassar}.
The mass fraction of $^{36}$Ar itself is observed to increase by a factor of $\sim$10 due to its lower measured capture-cross section.
Especially interesting is the $\sim$45\% decrease in the calculated mass fraction of the important cosmo/geo-chronometer $^{40}$K
implying a weaker contribution of the secondary $s$-process relative to primary production. As shown in Frank et al. \cite{Frank}, 
the mass fraction of $^{40}$K differs considerably over time whether it is primary only or secondary only. 
For example, with a larger primary production of $^{40}$K which is the dominant initial heat generator in Earth-like exoplanets, 
considerable heating would occur in these worlds even early in the Galaxy history.

The measurements of the $^{36}$Ar$(n,\gamma)$ cross sections affect also the calculation of the natural $^{37}$Ar background activity in the atmosphere, the interpretation of
$^{37}$Ar emission rates in underground nuclear explosion monitoring \cite{LLC2} and the investigation of atmospheric air circulation \cite{Loosli1}.
The detection of $^{37}$Ar by AMS demonstrated here opens the way to an alternative method for the monitoring of environmental samples \cite{AMS14}.
Similarly to $^{37}$Ar, the $^{38}$Ar$(n,\gamma)^{39}$Ar reaction contributes to the $^{39}$Ar production rate in the atmosphere \cite{Loosli1} and determines the initial value
for the use of $^{39}$Ar as a groundwater dating chronometer \cite{Corcho, Loosli2, Collon}.
\FloatBarrier
In summary, first measurements of the neutron capture cross sections of $^{36}$Ar and $^{38}$Ar at stellar energies were performed.
The experimental value for $^{36}$Ar, in particular, is smaller than the one adopted so far from theoretical calculations and evaluations by a factor of $\sim$10.
Nucleosynthesis calculations for the weak $s$-process regime using the measured cross sections are shown to increase the mass fraction of $^{36}$Ar by a factor of $\sim$10 and
lower the residual mass fraction of neutron-rich nuclides in the region A=36-48 by 10 to 50\%.
The $^{36,38}$Ar$(n,\gamma)$ cross sections affect the interpretation of environmental monitoring using $^{37}$Ar or $^{39}$Ar as geophysical tracers.

\begin{acknowledgments}
We would like to thank the SARAF and LiLiT (Soreq NRC) and the ATLAS operation staffs for their dedicated help during the experiments.
This work was supported in part by the Israel Science Foundation (Grant No. 1387/15), by the Pazy Foundation (Israel), the Israel Ministry of Science (Eshkol Grant No. 18145),
the US Department of Energy, Office of Nuclear Physics, under Award No. DE-AC02-06CH11357. D.S.G. acknowledges the support by the U.S. Department of Energy, Office of Nuclear Physics, under Award No. DE-FG02-96ER40978.
This research has received funding from the European Research Council under the European Unions's Seventh Framework Program (FP/2007-2013)/ERC Grant Agreement No. 615126.
\end{acknowledgments}



\newpage

\begin{center}
\textbf{{\huge Supplementary material}}
\end{center}

We present here supplementary material referred to in the paper.
\begin{figure}[h]
\centering
\includegraphics[width=0.5\columnwidth]{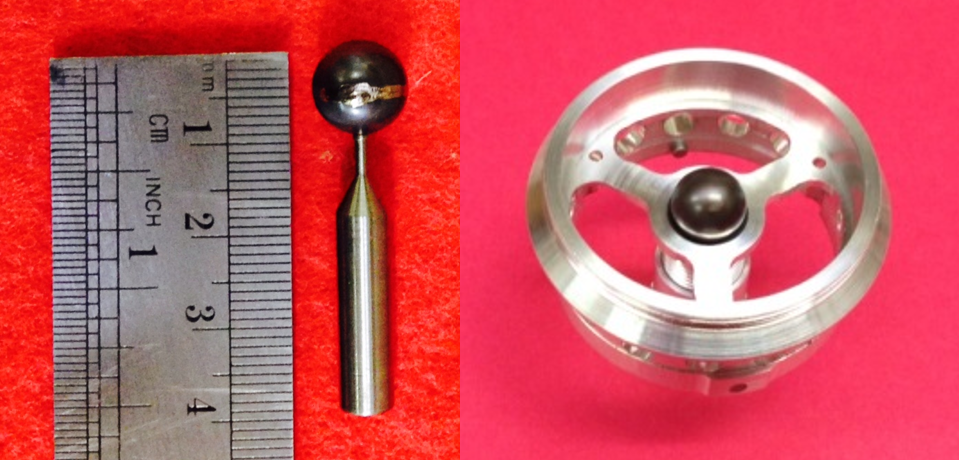}
  \caption{\label{fig:gas_sphere} Ti sphere used as container for the pressurized $^{A}$Ar gas for irradiation (left), and the sphere target holder for irradiation at SARAF-LiLIT (right).}
\end{figure}

\begin{table*}[h]
\centering
\caption{\label{table:samples_supp}Samples used and the results of the $^{A+1}$Ar/$^A$Ar ratio after the irradiation for all gas samples.
$^{36}$Ar and $^{38}$Ar gas samples \cite{enrich} were enriched to 99.935\% and 99.957\% for the respective isotopes.
The final $^{37}$Ar/$^{36}$Ar ratios were obtained by taking a weighted average of the AMS and LLC results (Fig. \ref{fig:AMS_LLC}).
Sphere \#52a was irradiated with 1 mm thick Cd shield to estimate the epithermal neutron fraction.}
\begin{ruledtabular}
\ra{1.3}
\begin{tabular}{l c c c c}
Sphere \# & $^A$Ar gas & $^A$Ar mass (mg) & Irradiation & $^{A+1}$Ar/$^A$Ar ratio\\ [0.5ex]
\hline
39 & $^{36}$Ar & 24.5 & LiLiT & $8.6(6)\times10^{-13}$\\
52a & $^{36}$Ar & 19.9 & reactor, 20 min., 80 kW, w/ Cd & $1.4(1)\times10^{-12}$\\
60 & $^{36}$Ar & 22.6 & reactor, 20 min., 80 kW, w/o Cd & $3.3(2)\times10^{-10}$\\ [0.5ex]
\hline
59 & $^{38}$Ar & 19.5 & LiLiT & $4.0(4)\times10^{-13}$\\
54 & $^{38,nat}$Ar & 12.8 ($^{38}$Ar/$^{nat}$Ar = 11.7) & reactor, 40 sec, 5 MW & $8.6(9)\times10^{-11}$\\
52b & $^{38,nat}$Ar & 15.8 ($^{38}$Ar/$^{nat}$Ar = 10.2) & reactor, 20 sec, 3 MW & $1.8(2)\times10^{-11}$\\ [0.5ex]
\end{tabular}
\end{ruledtabular}
\end{table*}

\begin{figure}[h]
\centering
\includegraphics[width=0.5\columnwidth]{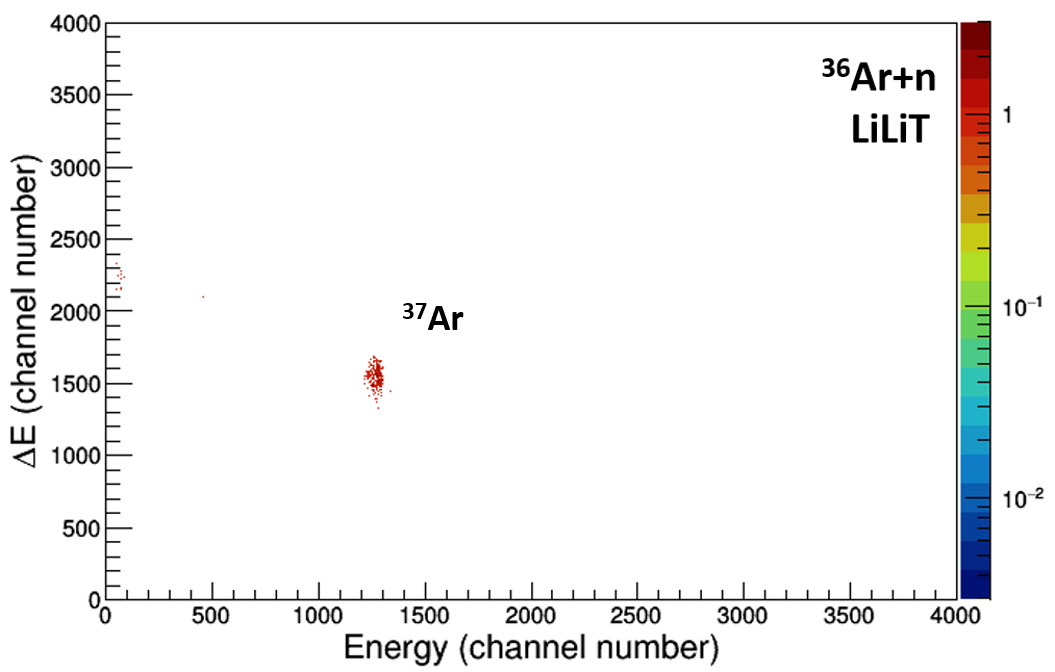}
  \caption{\label{fig:36_LiLiT}Identification spectra of the $^{37}$Ar counts with the $\Delta$E-E telescope of Si detectors for the LiLiT irradiated sphere \#39.}
\end{figure}

\begin{figure}[h]
\centering
\includegraphics[width=0.5\columnwidth]{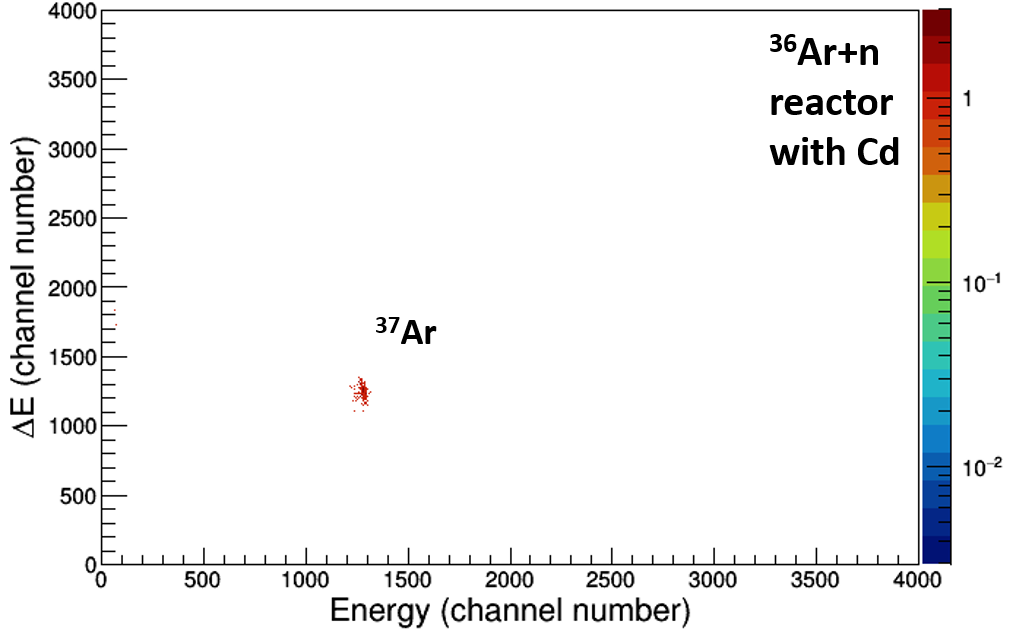}
  \caption{\label{fig:36_Cd}Same as Fig. \ref{fig:36_LiLiT} for the reactor irradiated sphere with Cd \#52a.}
\end{figure}

\begin{figure}[h]
\centering
\includegraphics[width=0.5\columnwidth]{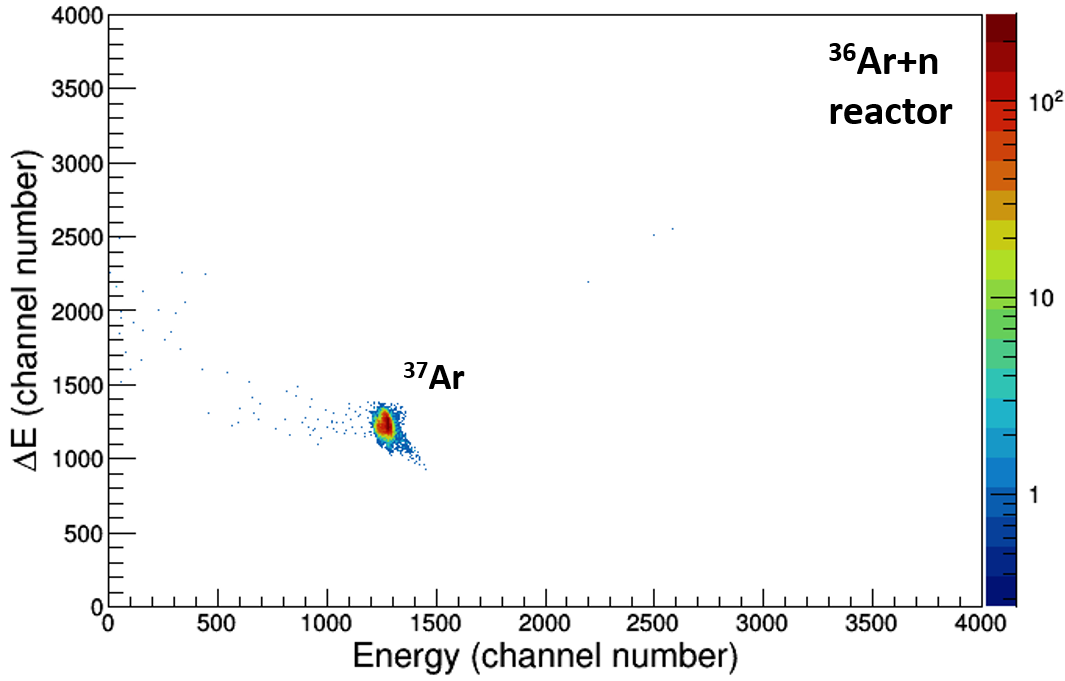}
  \caption{\label{fig:36_reactor}Same as Fig. \ref{fig:36_LiLiT} for the reactor irradiated sphere (without Cd) \#60.}
\end{figure}

\begin{figure}[h]
\centering
\includegraphics[width=0.5\columnwidth]{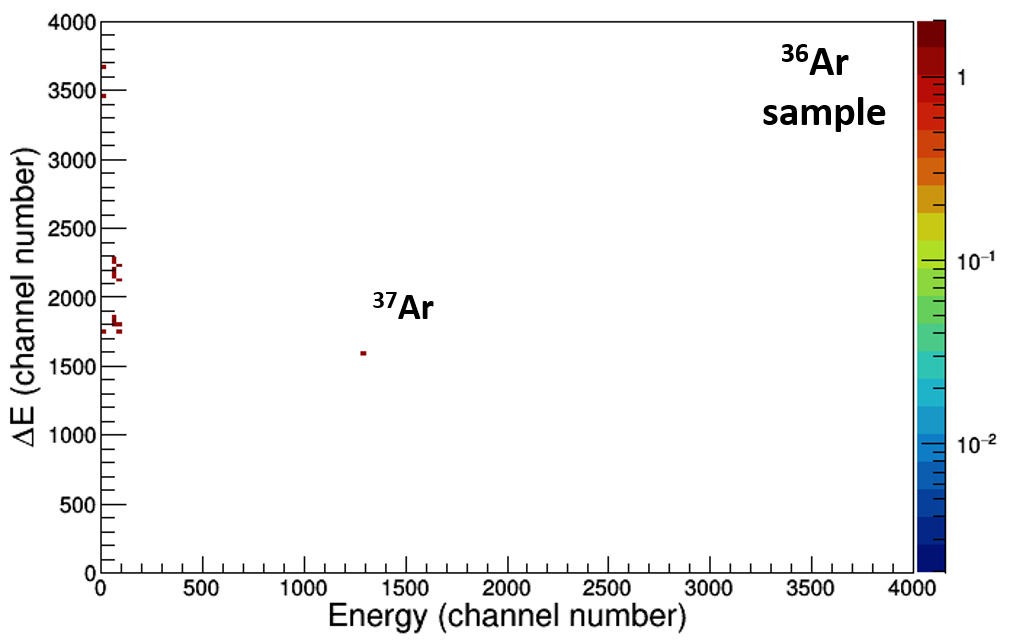}
  \caption{\label{fig:36_blank}Same as Fig. \ref{fig:36_LiLiT} for a non irradiated $^{36}$Ar (blank). One $^{37}$Ar count detected for this sample over 6.5 hours,
  likely due to a memory effect in the ion source, corresponds to a concentration $^{37}$Ar/$^{36}$Ar=$9\times10^{-16}$.}
\end{figure}

\begin{figure}[h]
\centering
\includegraphics[width=0.99\columnwidth]{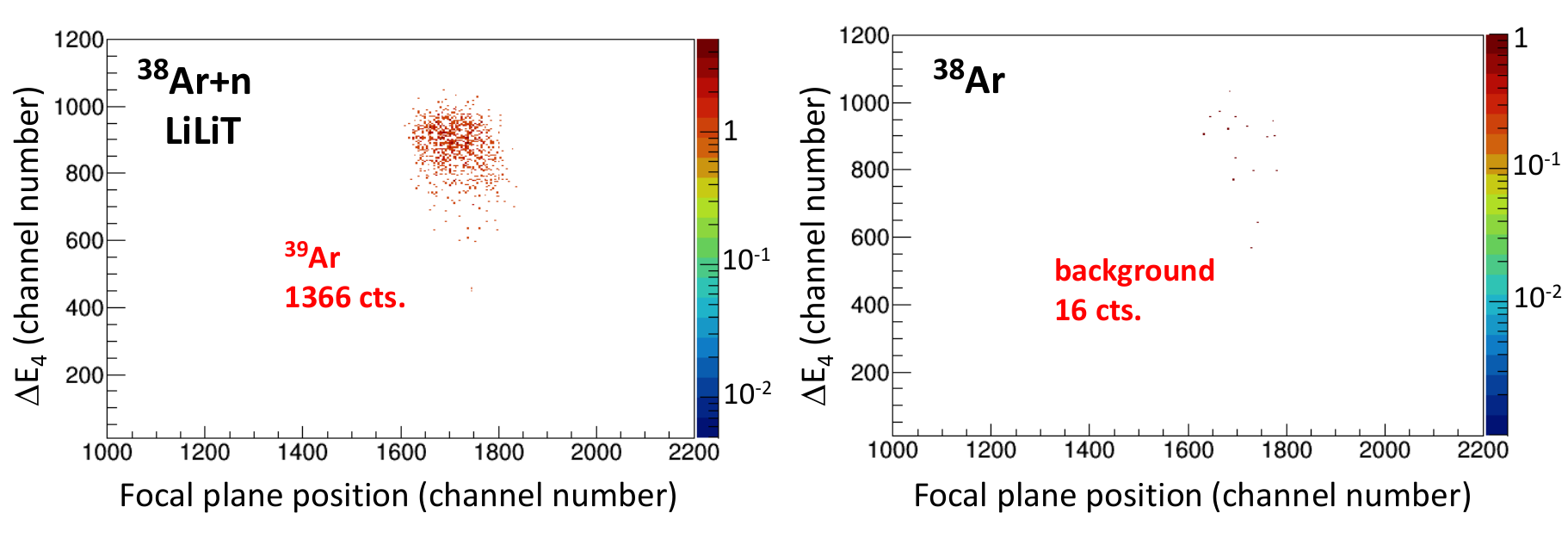}
  \caption{\label{fig:2d_gate} Two-dimensional spectra of $\Delta$E$_4$ vs. focal plane position gated on the $^{39}$Ar region-of-interest in the time-of-flight, $\Delta$E$_2$ and $\Delta$E$_3$ detector parameters.}
\end{figure}

\begin{table}[h]
\centering
\caption{\label{table:39}AMS results obtained for the LiLiT irradiated sphere (\#39) $^{37}$Ar/$^{36}$Ar ratio. Statistical uncertainties are given in (). An additional systematic uncertainty
of 10\% is due to the $^{37}$Ar transmission.}
\begin{ruledtabular}
\ra{1.3}
\begin{tabular}{l c c c c c}
run \# & $N(37)$ & time (sec) & $i_{36}$ (nA) & $^{37}$Ar/$^{36}$Ar ratio\\ [0.5ex]
\hline
40 & 94(10) & 5207.8 & 1.66(1) & $9.0(10)\times10^{-13}$\\
41+42 & 68(8) & 4333.8 & 1.6(3) & $8.0(15)\times10^{-13}$\\
43 & 37(6) & 3866.7 & 1.1(1) & $7.2(13)\times10^{-13}$\\
44 & 50(7) & 3918.6 & 1.00(5) & $1.1(2)\times10^{-12}$\\
45 & 55(7) & 3489.1 & 1.50(6) & $8.8(13)\times10^{-13}$\\
\hline
weighted &  & & & \multirow{2}{*}{$8.8(5)\times10^{-13}$} \\
average &&&& \\ 
\bottomrule
\bottomrule
\end{tabular}
\end{ruledtabular}
\end{table}

\begin{figure}[h]
\centering
\includegraphics[width=0.6\columnwidth]{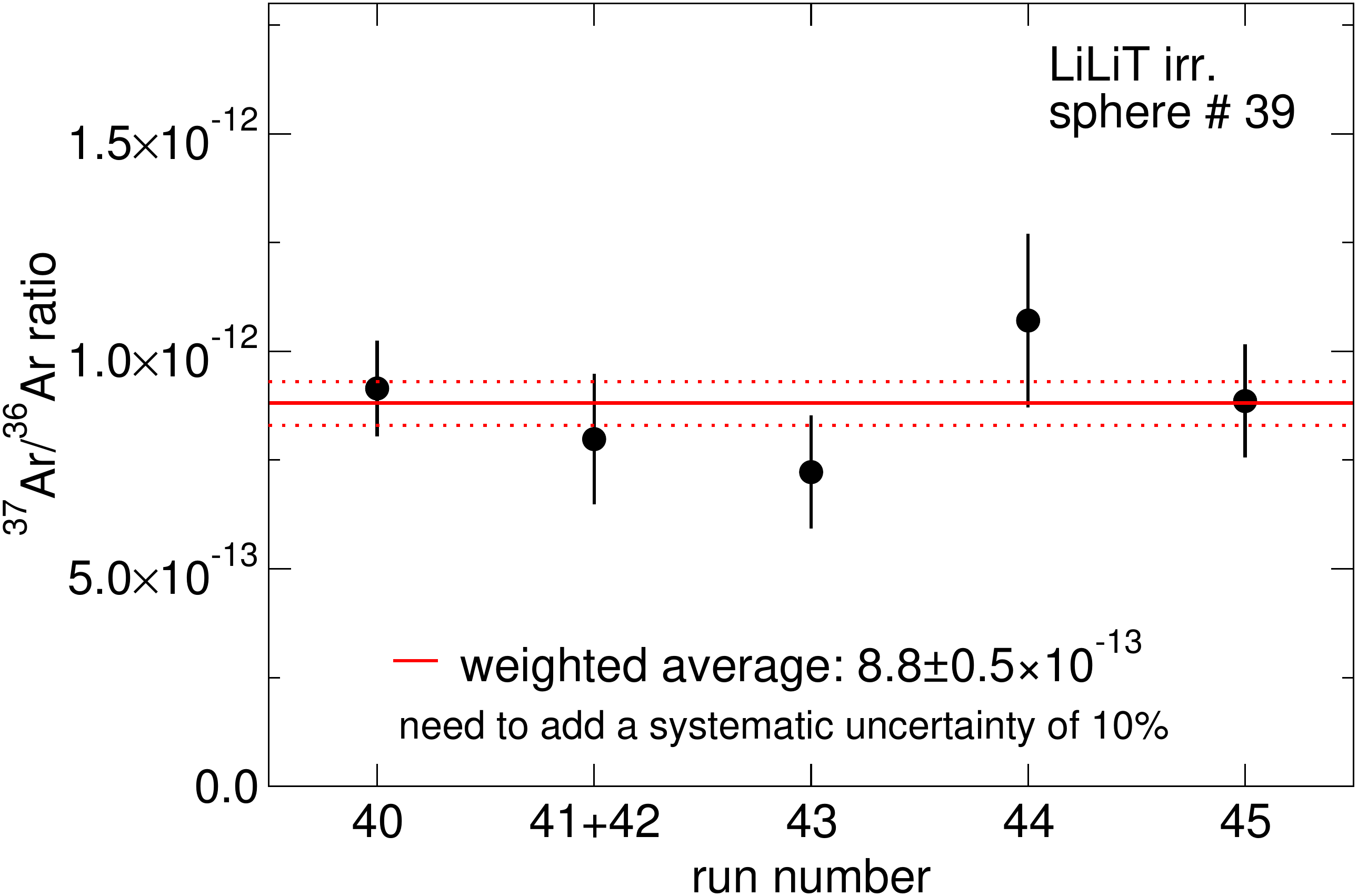}
  \caption{\label{fig:39}Repeated measurements of the $^{37}$Ar/$^{36}$Ar ratio for the LiLiT irradiated sphere (\#39). See Table \ref{table:39} for numerical results.}
\end{figure}

\begin{table}[h]
\centering
\caption{\label{table:52b}Same as Table \ref{table:39} for the Cd shielded sphere irradiated at the reactor (\#52a).}
\begin{ruledtabular}
\ra{1.3}
\begin{tabular}{l c c c c c}
run \# & $N(37)$ & time (sec) & $i_{36}$ (nA) & $^{37}$Ar/$^{36}$Ar ratio\\ [0.5ex]
\hline
52 & 48(7) & 2526.3 & 1.4(4) & $1.0(3)\times10^{-12}$\\
53 & 122(11) & 3135.3 & 1.9(1) & $1.4(1)\times10^{-12}$\\
\hline
weighted  &  & & & \multirow{2}{*}{$1.3(1)\times10^{-12}$}\\
average \\
\end{tabular}
\end{ruledtabular}
\end{table}

\begin{table}[h]
\centering
\caption{\label{table:60}Same as Table \ref{table:39} for the sphere irradiated at the reactor (\#60).}
\begin{ruledtabular}
\ra{1.3}
\begin{tabular}{l c c c c c}
run \# & $N(37)$ & time (sec) & $i_{36}$ (nA) & $^{37}$Ar/$^{36}$Ar ratio\\ [0.5ex]
\hline
54 & 10990(105) & 579.2 & 2.7(1) & $4.8(2)\times10^{-10}$\\
55 & 9228(96) & 563.2 & 2.85(3) & $3.95(6)\times10^{-10}$\\
56 & 8574(93) & 566.6 & 2.79(3) & $3.72(6)\times10^{-10}$\\
57 & 8595(93) & 567.5 & 2.86(7) & $3.6(1)\times10^{-10}$\\
58 & 7293(85) & 564.8 & 2.7(2) & $3.2(2)\times10^{-10}$\\
59 & 8942(95) & 644.9 & 2.81(2) & $3.38(4)\times10^{-10}$\\
60 & 9402(97) & 645.5 & 2.89(7) & $3.44(8)\times10^{-10}$\\
\hline
weighted &  & & & \multirow{2}{*}{$3.61(2)\times10^{-10}$}\\
average \\
\bottomrule
\bottomrule
\end{tabular}
\end{ruledtabular}
\end{table}
\begin{figure}[h]
\centering
\includegraphics[width=0.6\columnwidth]{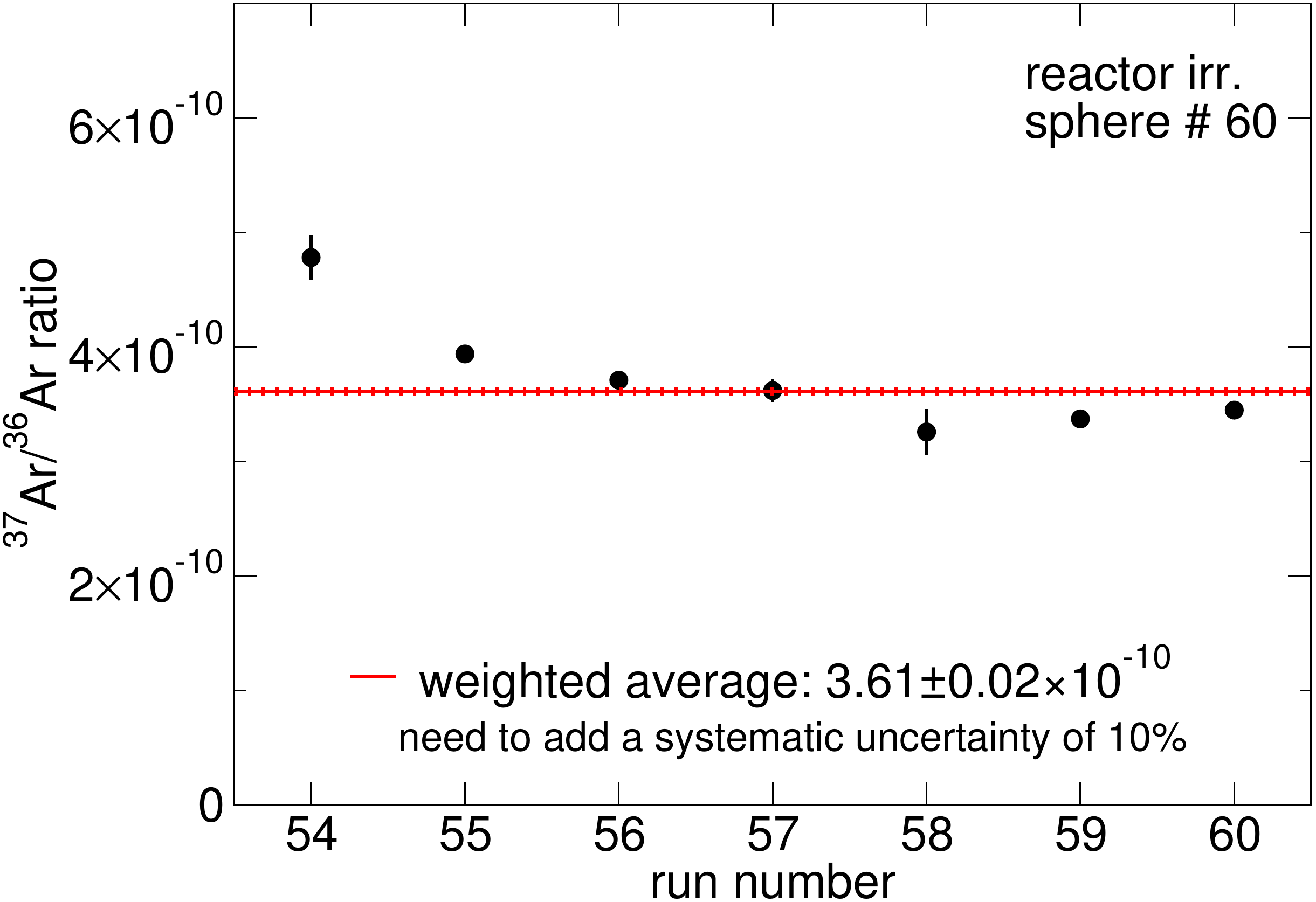}
  \caption{\label{fig:60}Same as Fig. \ref{fig:39} for the sphere irradiated at the reactor (\#60). See Table \ref{table:60} for numerical results.}
\end{figure}
\begin{figure}[h]
\centering
\includegraphics[trim={0 3.2cm 0 1.6cm},clip,width=0.6\columnwidth]{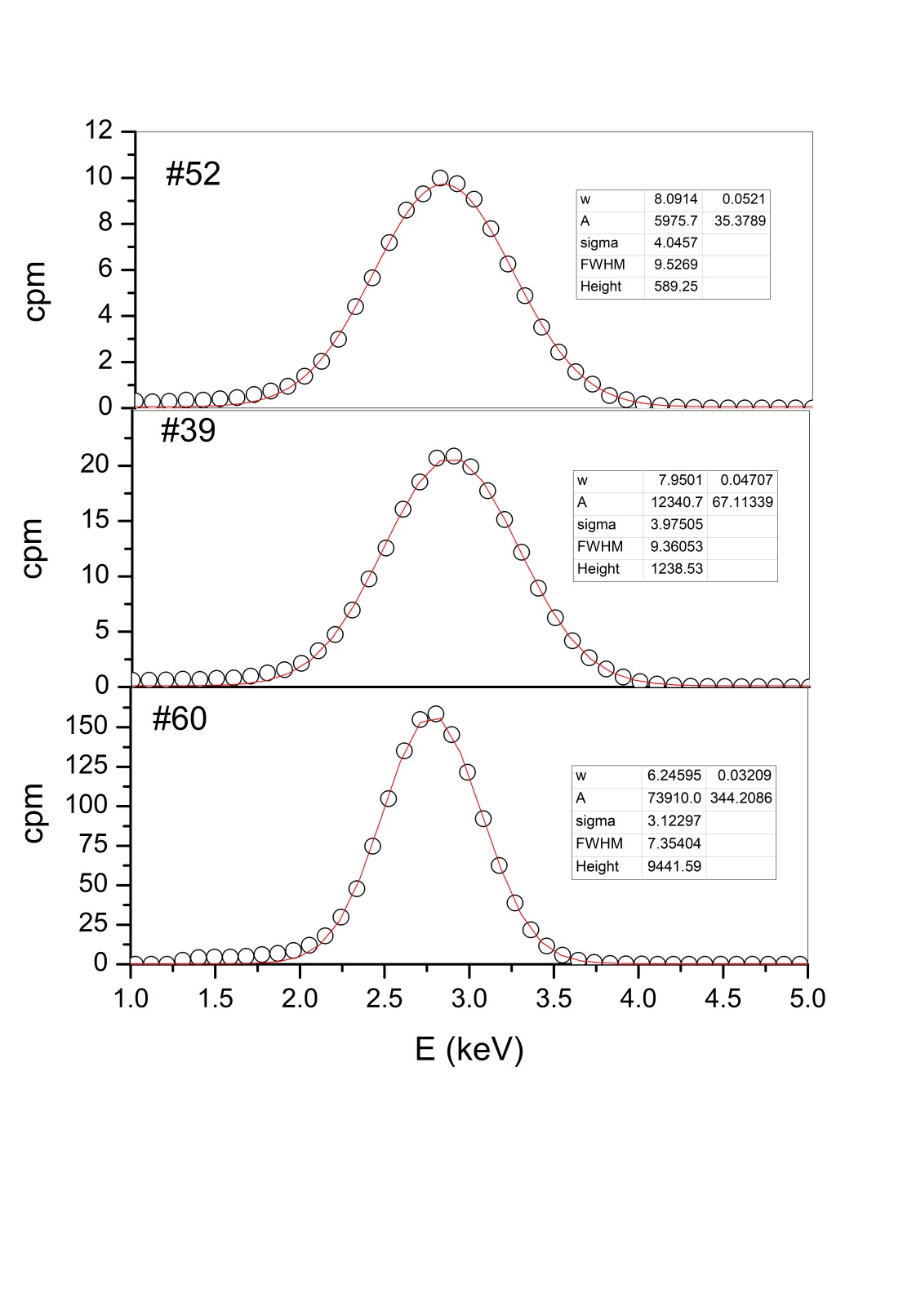}
  \caption{\label{fig:LLC} Energy spectra (in counts per minute) and gauss fit of $^{37}$Ar measurements by low-level counting (LLC).
  The peak energy correspond to the 2.82 keV K-capture decay energy of $^{37}$Ar.}
\end{figure}
\begin{table}[h]
\centering
\caption{\label{table:LLC}LLC results of the $^{37}$Ar/$^{36}$Ar ratios for all three samples.}
\begin{ruledtabular}
\ra{1.3}
\begin{tabular}{l c}
sample & $^{37}$Ar/$^{36}$Ar ratio\\ [0.5ex]
\hline
LiLiT & $8.59(70)\times10^{-13}$\\
reactor with Cd & $1.44(12)\times10^{-12}$\\
reactor (w/o Cd) & $3.23(25)\times10^{-10}$\\
\end{tabular}
\end{ruledtabular}
\end{table}

\begin{figure}[h]
\centering
\includegraphics[width=0.6\columnwidth]{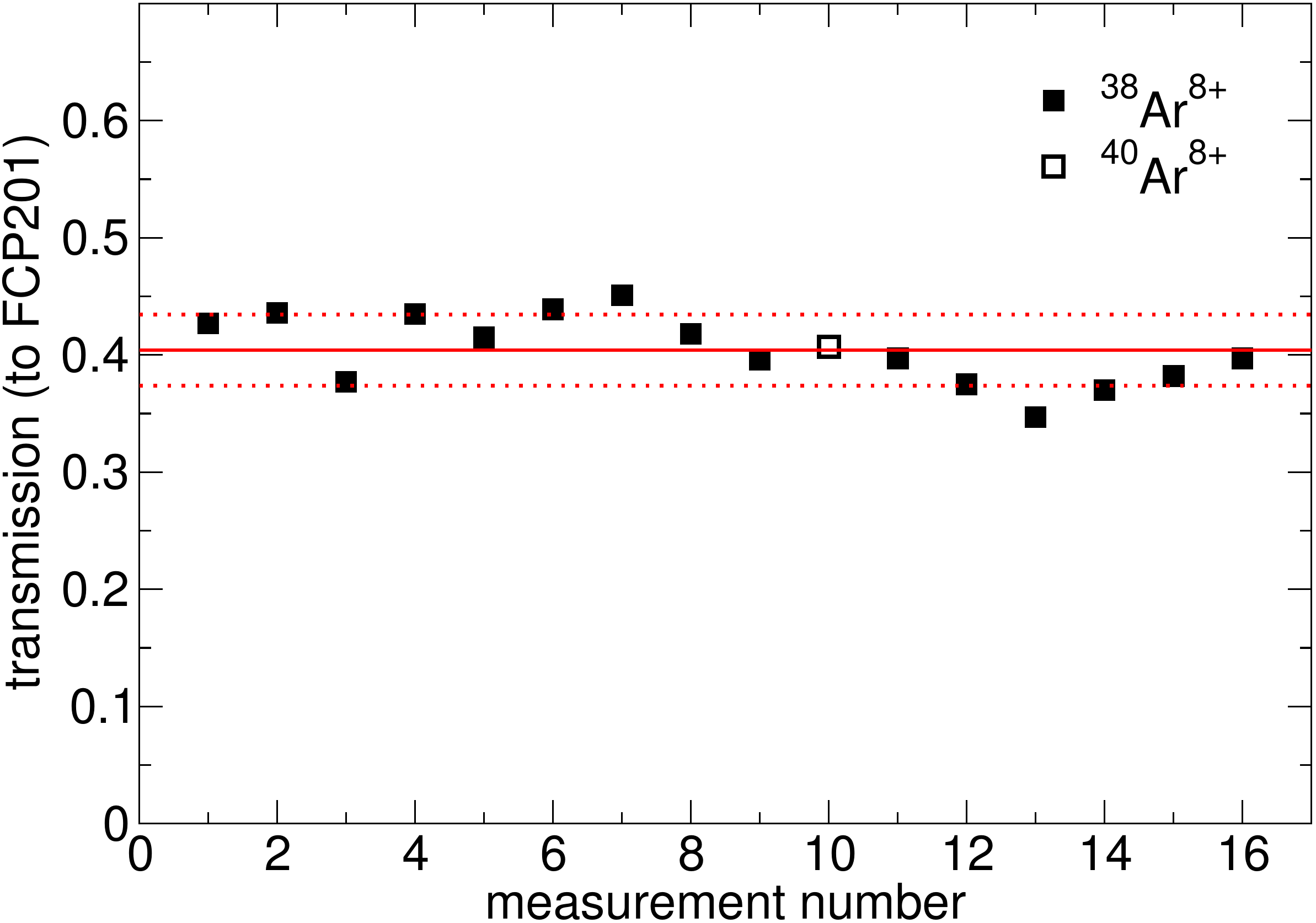}
  \caption{\label{fig:trans}Transmission efficiency of $^{38}$Ar$^{8+}$ and $^{38}$Ar$^{8+}$.}
\end{figure}

\begin{table}[h]
\centering
\begin{threeparttable}
\centering
\caption{\label{table:59}Analysis of the LiLiT irradiated sphere (\# 59) $^{39}$Ar/$^{38}$Ar ratio.}
\begin{ruledtabular}
\ra{1.3}
\begin{tabular}{l c c c c c}
run \# & $N(39)$ & time (sec) & $i_{38}$ ($\mu$A) & $^{39}$Ar/$^{38}$Ar ratio\\ [0.5ex]
\hline
53 & 680(19) & 3584 & 0.69(7) & $3.9(4)\times10^{-13}$\\
54 & 589(20) & 3606 & 0.48(5) & $4.8(5)\times10^{-13}$\\
55 & 756(19) & 3855 & 0.71(7) & $3.9(4)\times10^{-13}$\\
56 & 610(17) & 3584 & 0.56(6) & $4.2(5)\times10^{-13}$\\
57 & 656(18) & 3538 & 0.71(7) & $3.7(4)\times10^{-13}$\\
63 & 659(18) & 3653 & 0.60(6) & $4.3(5)\times10^{-13}$\\
69 & 783(23) & 3697 & 0.64(6) & $4.7(5)\times10^{-13}$\\
70 & 791(23) & 3698 & 0.65(7) & $4.6(5)\times10^{-13}$\\
73 & 374(14) & 1826 & 0.63(6) & $4.6(5)\times10^{-13}$\\
\hline
weighted &  & & & \multirow{2}{*}{$4.2(2)\times10^{-13}$}\\
average \\
\hline
blank  &  & & & $2.1(2)\times10^{-14}$\\
\hline
final ratio  &  & & & $4.0(4)\times10^{-13}$ \tnote{a}\\
\end{tabular}
\end{ruledtabular}
\begin{tablenotes}
            \item[a] including the 3.3\% uncertainty of the detector efficiency.
 \end{tablenotes}
\end{threeparttable}
\end{table}
\begin{figure}[h]
\centering
\includegraphics[width=0.6\columnwidth]{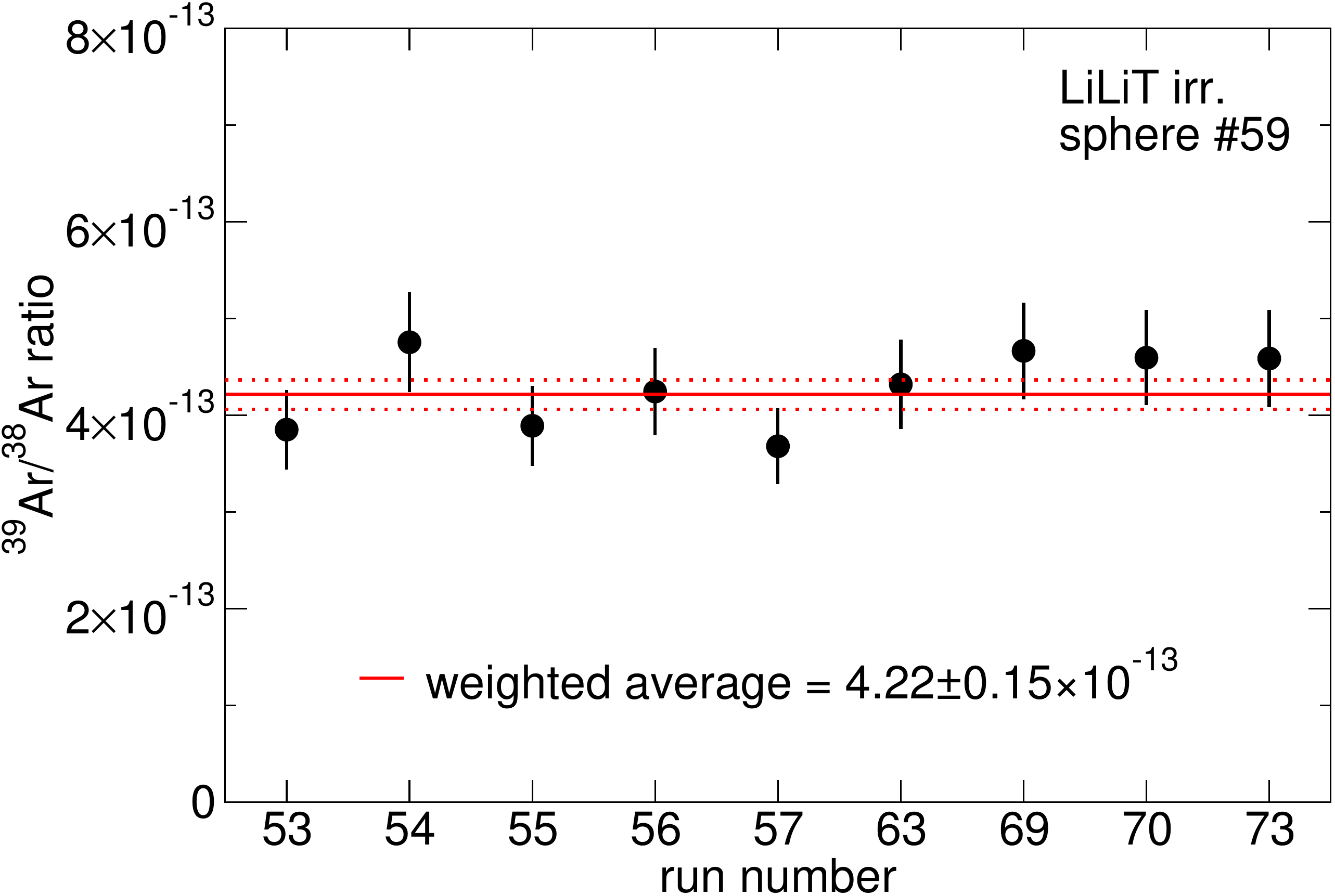}
  \caption{\label{fig:59}Repeated measurements of the $^{39}$Ar/$^{38}$Ar ratio for the LiLiT irradiated sphere (\#59). See Table \ref{table:59} for numerical results.}
\end{figure}
\begin{table}[h]
\centering
\begin{threeparttable}
\centering
\caption{\label{table:54}AMS results obtained for the reactor irradiated sphere (54) $^{39}$Ar/$^{38}$Ar ratio.}
\begin{ruledtabular}
\ra{1.3}
\begin{tabular}{l c c c c c}
run \# & $N(39)$ & time (sec) & $i_{38}$ ($\mu$A) & $^{39}$Ar/$^{38}$Ar ratio\\ [0.5ex]
\hline
76 & 91629(1392) & 3628 & 0.42(2) & $8.5(9)\times10^{-11}$\\
77 & 65234(1081) & 3606 & 0.28(3) & $9.2(10)\times10^{-11}$\\
78 & 76924(1236) & 3102 & 0.39(4) & $8.9(9)\times10^{-11}$\\
79 & 115416(1847) & 3601 & 0.53(5) & $8.4(9)\times10^{-11}$\\
80 & 103310(1616) & 3542 & 0.48(5) & $8.6(9)\times10^{-11}$\\
81 & 119096(1939) & 3381 & 0.58(6) & $8.6(9)\times10^{-11}$\\
82 & 124323(1979) & 3545 & 0.58(6) & $8.5(9)\times10^{-11}$\\
83 & 119768(1980) & 3580 & 0.55(6) & $8.6(9)\times10^{-11}$\\
84 & 115483(1880) & 3587 & 0.53(5) & $8.5(9)\times10^{-11}$\\
\hline
weighted &  & & & \multirow{2}{*}{$8.6(3)\times10^{-11}$}\\
average \\
\hline
blank  &  & & & $2.1(2)\times10^{-14}$\\ 
\hline
final ratio  &  & & & $8.6(9)\times10^{-11}$ \tnote{a}\\
\end{tabular}
\end{ruledtabular}
\begin{tablenotes}
            \item[a] including the 3.3\% uncertainty of the detector efficiency.
 \end{tablenotes}
\end{threeparttable}
\end{table}
\begin{figure}[h]
\centering
\includegraphics[width=0.6\columnwidth]{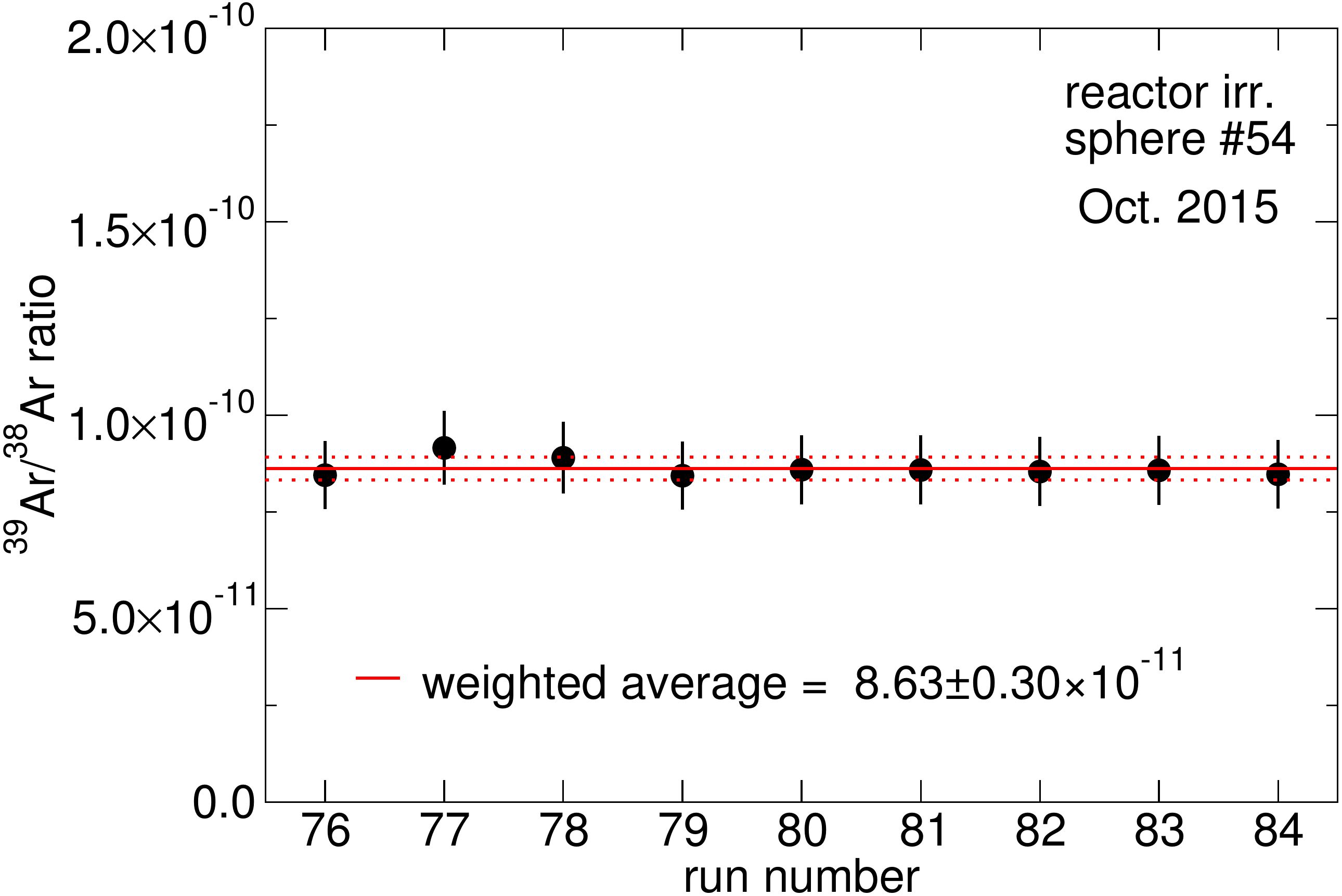}
  \caption{\label{fig:54}Same as Fig. \ref{fig:59} for the reactor irradiated sphere (\#54). See Table \ref{table:54} for numerical results.}
\end{figure}
\begin{table}[h]
\centering
\begin{threeparttable}
\centering
\caption{\label{table:52b}Same as Table \ref{table:54} for the reactor irradiated sphere (\#52b).}
\begin{ruledtabular}
\ra{1.3}
\begin{tabular}{l c c c c c}
run \# & $N(39)$ & time (sec) & $i_{38}$ ($\mu$A) & $^{39}$Ar/$^{38}$Ar ratio\\ [0.5ex]
\hline
157 & 3597(47) & 900 & 0.30(2) & $1.84(14)\times10^{-11}$\\
158 & 3349(56) & 908 & 0.25(2) & $2.05(16)\times10^{-11}$\\
159 & 3192(46) & 901 & 0.28(2) & $1.77(14)\times10^{-11}$\\
161 & 3180(53) & 900 & 0.27(2) & $1.84(14)\times10^{-11}$\\
163 & 6674(110) & 1231 & 0.44(3) & $1.74(13)\times10^{-11}$\\
164 & 11207(259) & 2037 & 0.42(3) & $1.84(14)\times10^{-11}$\\
\hline
weighted &  & & & \multirow{2}{*}{$1.84(6)\times10^{-11}$}\\
average \\
\hline
blank  &  & & & $1.37(17)\times10^{-13}$\\
\hline
final ratio  &  & & & $1.82(15)\times10^{-11}$ \tnote{a}\\
\end{tabular}
\end{ruledtabular}
\begin{tablenotes}
            \item[a] including the 3.3\% uncertainty of the detector efficiency.
 \end{tablenotes}
\end{threeparttable}
\end{table}
\begin{figure}[h]
\centering
\includegraphics[width=0.6\columnwidth]{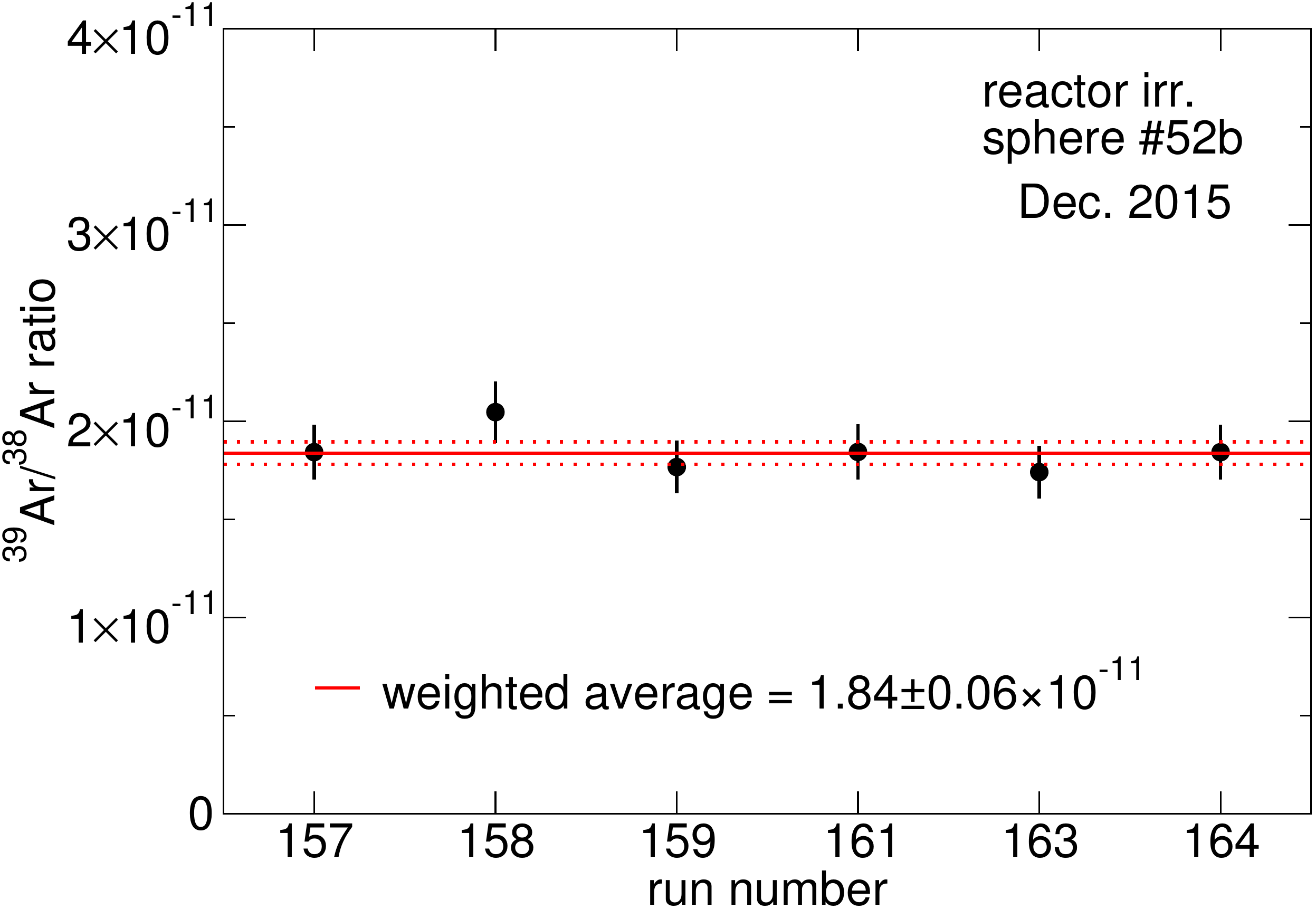}
  \caption{\label{fig:52b}Same as Fig. \ref{fig:59} for the reactor irradiated sphere (\#52b). See Table \ref{table:52b} for numerical results.}
\end{figure}
\begin{figure}[h]
\centering
\includegraphics[width=0.6\columnwidth]{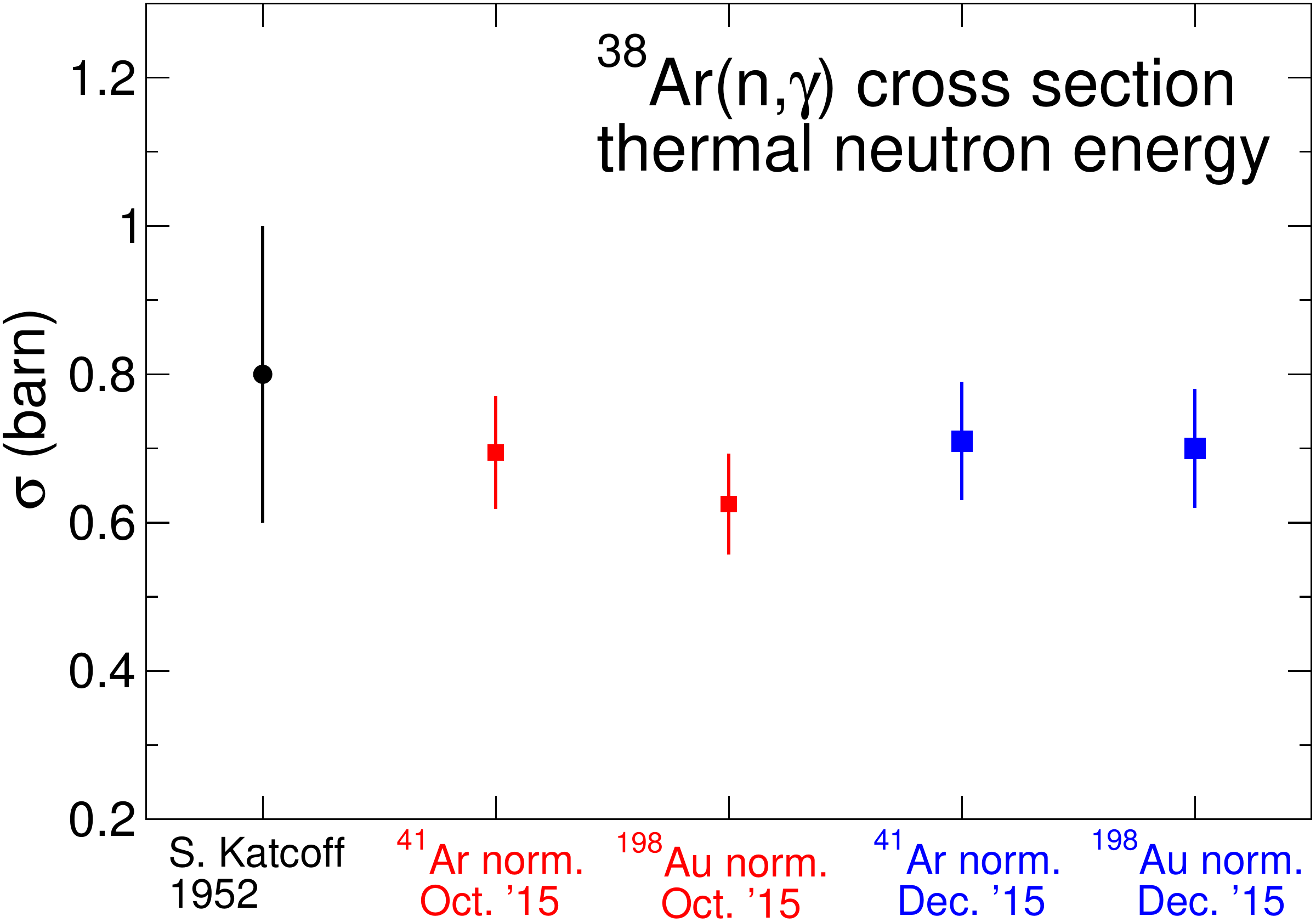}
  \caption{\label{fig:thermal} Comparison of the measured  $^{38}$Ar thermal cross section using external $^{198}$Au and internal $^{41}$Ar neutron monitoring.}
\end{figure}
\begin{figure}[h]
\centering
\includegraphics[width=0.99\columnwidth]{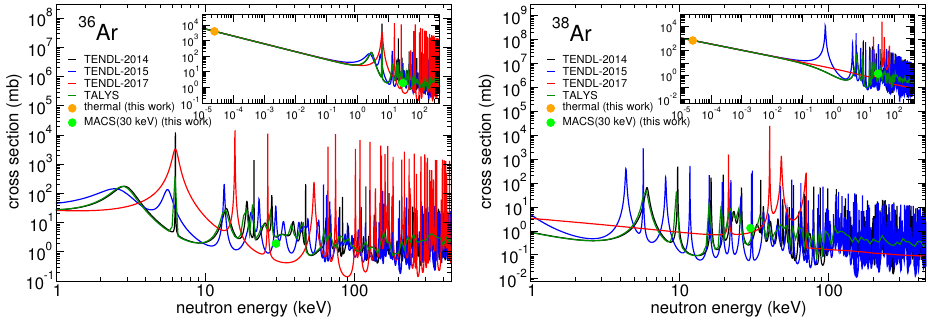}
  \caption{\label{fig:exc_func} Excitation functions of the $^{36,38}$Ar$(n,\gamma)$ reaction calculated with the Hauser-Feshbach codes indicated.
  The experimental values of the thermal cross section and of the MACS(30 keV) determined in this work are shown.
  The excitation functions are matched to the thermal cross sections measured in this work.}
\end{figure}

\begin{figure}[h]
\centering
\includegraphics[width=0.6\columnwidth]{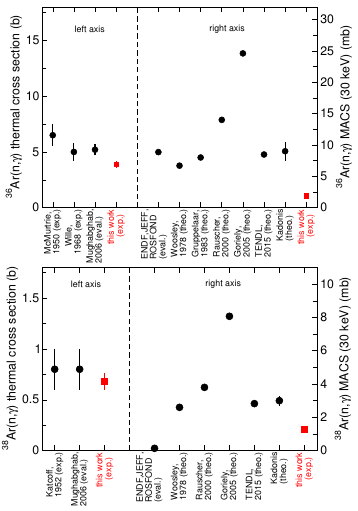}
  \caption{\label{fig:36_38Ar_TENDL} Comparison of the $^{36}$Ar (top) and $^{38}$Ar (bottom) thermal and Maxwellian Average (30 keV) cross sections measured in this work (red squares)
  to measured (thermal) and theoretical data (black circles). Numerical values are listed in Table \ref{table:MACS_comp} (main paper).}
\end{figure}
\FloatBarrier

\end{document}